\documentclass{emulateapj}
\newcommand{\as}[2]{$#1''\,\hspace{-1.7mm}.\hspace{.1mm}#2$}

\newcommand{\HII}{\mbox{H{\sc ii}}}
\newcommand{\NII}{\mbox{[N{\sc ii}]}}
\newcommand{\OI}{\mbox{[O{\sc i}]}}
\newcommand{\OIII}{\mbox{[O{\sc iii}]}}
\newcommand{\SII}{\mbox{[S{\sc ii}]}}
\def\approxlt{\lower.2em\hbox{$\buildrel < \over \sim$}}
\def\approxgt{\lower.2em\hbox{$\buildrel > \over \sim$}}

\def\gtrsim{\mathrel{\hbox{\rlap{\hbox{\lower4pt\hbox{$\sim$}}}\hbox{$>$}}}}

\def\lesssim{\mathrel{\hbox{\rlap{\hbox{\lower4pt\hbox{$\sim$}}}\hbox{$<$}}}}

\def\la{\mathrel{\hbox{\rlap{\hbox{\lower4pt\hbox{$\sim$}}}\hbox{$<$}}}}
\def\ga{\mathrel{\hbox{\rlap{\hbox{\lower4pt\hbox{$\sim$}}}\hbox{$>$}}}}

\received{}
\revised{}
\accepted{}

\slugcomment{\date}

\shorttitle{Intensely Star-Forming Galaxies at Redshift 2}
\shortauthors{Lehnert et al.}

\begin{document}

\title{Physical conditions in the ISM of intensely star-forming galaxies
at redshift$\sim$2\altaffilmark{1}}

\author{M. D. Lehnert, N. P. H. Nesvadba\altaffilmark{2}, L. Le Tiran,
P. Di Matteo, W. van Driel, L.~S. Douglas, L. Chemin}

\affil{GEPI, Observatoire de Paris, CNRS, Universit\'e Paris Diderot,
5 place Jules Janssen, 92190 Meudon, France}

\and

\author{F. Bournaud}
\affil{CEA, IRFU, SAp, 91191 Gif-sur-Yvette, France\\ Laboratoire AIM,
CEA-Saclay - CNRS - Universit\'e Paris Diderot, Saclay, France }

\altaffiltext{1}{Data obtained as part of Programme IDs 075.A-0466,
076.A-0527, 077.A-0576, 078.A-0600, and 079.A-0341 at the ESO-VLT and
based on observations made with the NASA/ESA Hubble Space Telescope,
obtained from the data archive at the Space Telescope Science Institute,
which is operated by the association of universities for research in
astronomy, inc., under NASA contract NAS 5-26555.}

\altaffiltext{2}{{\it Present Address:} Institut d'Astrophysique Spatiale,
UMR 8617, CNRS, Universit\'e Paris-Sud, B\^atiment 121, F-91405 Orsay
Cedex, France}

\begin{abstract}
We analyze the physical conditions in the interstellar gas of 11 actively
star-forming galaxies at z$\sim$2, based on integral-field spectroscopy
from the ESO-VLT and HST/NICMOS imaging. We concentrate on the high
H$\alpha$ surface brightnesses, large line widths, line ratios and
the clumpy nature of these galaxies.  We show that photoionization
calculations and emission line diagnostics imply gas pressures and
densities that are similar to the most intense nearby star-forming regions
at z=0 but over much larger scales (10-20 kpc).  A relationship between
surface brightness and velocity dispersion can be explained through
simple energy injection arguments and a scaling set by nearby galaxies
with no free parameters.  The high velocity dispersions are a natural
consequence of intense star formation thus regions of high velocity
dispersion are not evidence for mass concentrations such as bulges or
rings.  External mechanisms like cosmological gas accretion generally
do not have enough energy to sustain the high velocity dispersions.
In some cases, the high pressures and low gas metallicites may make it
difficult to robustly distinguish between AGN ionization cones and star
formation, as we show for BzK-15504 at z$=$2.38.  We construct a picture
where the early stages of galaxy evolution are driven by self-gravity
which powers strong turbulence until the velocity dispersion is high.
Then massive, dense, gas-rich clumps collapse, triggering star formation
with high efficiencies and intensities as observed.  At this stage,
the intense star formation is likely self-regulated by the mechanical
energy output of massive stars.
\end{abstract}


\section{Introduction}

Elucidating the physical processes that regulate global star
formation is one of the keys to understanding galaxy evolution.
The power-law relation, over several orders of magnitude, between
the warm HI and cold molecular gas surface density and star formation
intensity (``Schmidt-Kennicutt law'') is telling us that there must be
underlying physical processes that control and regulate star formation
\cite[e.g.,][]{kennicutt07}.  However, with a myriad of possible
mechanisms for regulating star formation over large scales -- cloud
formation and destruction, mechanical energy output from stars and AGN,
spiral density waves, turbulence induced by gravity and mechanical energy,
magnetic fields, mixing layers, and many others -- it is challenging
to distill a unifying explanation over many orders of magnitudes in
gas density.  One possible way of advancing our understanding of the
processes that regulate global star formation is by studying the most
extreme star-forming galaxies locally and at cosmological distances
\citep[e.g.,][]{lehnert96a, shapley03, verma07}.

Thanks to recent technological improvements, we are now able to study
the basic physical processes driving galaxy assembly directly and over
large ranges of cosmic time. In particular, the number of galaxies at
redshifts of 2$-$3 with detailed integral-field spectroscopic studies
in the rest-frame optical is still small, less than a few dozen,
but growing rapidly. These observations allow us to trace spatially
resolved emission-line properties, and to investigate galaxy kinematics
and the physical conditions of the ionized gas. Up to now, most studies
have focused on the kinematic properties of the galaxies and less so on
the properties of the emission lines themselves beyond their relative
velocities and widths. Recombination line fluxes were used to estimate
star-formation rates using simple prescriptions developed for low-redshift
galaxies \citep[][]{genzel06, fs06, vanstarkenburg08, law07, wright07,
nesvadba08a}.

With these estimates and assumptions, far-reaching conclusions have
been made regarding the modes of galaxy assembly and the drivers
of star formation in the early Universe. Many of these studies
\citep[e.g.,][]{genzel06, fs06} favor a scenario where high-redshift
galaxies exhibit gaseous disks of $\sim$10-20 kpc in size, which,
unlike galaxies at low redshift, are dynamically hot with large gas
velocity dispersions, $\sigma$, relative to their bulk velocities,
$v$ \citep[see also ][]{nesvadba06a}. The ratios of random to large
scale velocity shear at high redshift are of order v/$\sigma\sim$few,
compared to v$/\sigma \ge 10$ at low redshift. To explain these
observations, \citet{keres05,keres08,ocvirk08, dekel08} proposed a
scenario where galaxies at z$\sim 2$ accrete significant amounts of
cold gas, which after accumulating and forming gaseous disks, will
become gravitationally unstable and lead to the observed high star
formation rates \citep[][]{genzel08, dekel08}. In this picture, mergers
and hydrodynamic processes like feedback from star formation and AGN
play only a minor role in assembling early galaxies, beginning to play
a significant role only as a consequence of the build-up of the bulk of
the stellar mass in bulges and disks.

Given the faintness of the targets and the often low number of
diagnostic optical emission lines, usually only studying H$\alpha$ or
[OIII]$\lambda$5007, many of these results must rely on the assumption
that the gas conditions in the interstellar medium will overall be largely
similar to those in galaxies at low redshift. This assumption has not
been tested directly on the observed properties of high-redshift galaxies.

We already know of (or may plausibly expect) several major differences
between galaxies at high and at low redshift, which may strongly influence
the state of their interstellar medium and thus, their rest-frame
optical line emission. First, many galaxies at high redshift are less
evolved, with higher fractions \citep[][]{erb06c} of lower-metallicity gas
\citep[][]{erb06a, maiolino08}. This may have rather subtle observational
consequences. For example, \citet{robertson08} discuss the expected
morphology of an advanced (or mostly relaxed) merger of two gas-rich disk
galaxies as would be observed using state-of-the-art Integral-Field Units
(IFUs). They find that their model reproduces the observed properties
of high redshift galaxies as well as pure (and isolated) disk models,
which are often favored by observers in contrast to on-going or advanced
mergers \citep[][]{nesvadba06a, genzel06, genzel08, wright08}.

Second, virtually all high-redshift galaxies studied in detail
with IFUs have very high surface brightnesses of the recombination
lines. Given the impact of cosmological surface brightness dimming,
which is a strong function of redshift \citep[$\propto (z+1)^4$
for bolometric luminosities and spectral lines, $\propto (z+1)^5$
for broadband photometry due to the additional 'stretching' of the
continuum; see, e.g.,][]{tolman30}, and current observational limits,
all observations at high redshift will naturally be biased towards the
most luminous emission-line regions, and the most luminous galaxies. As
a result, all galaxies so far studied with IFUs have star-formation
rates of several 10s of M$_{\sun}$ yr$^{-1}$ or more, and typical
star-formation intensities, $SFI$, well above the critical threshold of
SFI$_{crit}\sim$ 0.1 M$_{\sun}$ yr$^{-1}$ kpc$^{-2}$\citep[][]{heckman03}
necessary to drive vigorous outflows in local galaxies. Galaxies with
$SFI \ga SFI_{crit}$ are observed to create and maintain strongly
over-pressurized bubbles of hot gas from the thermalized ejecta of
supernovae and massive young stars, which expand perpendicular to the
disk plane and produce galactic-scale outflows \citep[][]{heckman90,
lehnert96a, lehnert96b}. These outflows may play an important role
in rendering a starburst ``self-regulating'' \citep[][]{dopita94,
silk97}. The observational signatures of starburst-driven winds, such as
characteristic velocity offsets between the rest-frame UV absorption lines
and the systemic velocity \citep[][]{pettini00, erb04, erb06a}, or blue
wings in rest-frame optical emission lines \citep[e.g.][]{nesvadba07},
are commonly observed at high redshift. In a detailed study of a massive,
z$\sim 2.6$ starburst and the related outflow of a submillimeter-selected
galaxy at z$\sim 2.6$, \citet{nesvadba07} found that overall, the physical
properties of maximal starbursts at z$\sim$2$-$3 appear very similar
to local starbursts. The starburst appears to be self-regulating,
and characteristic, density-sensitive line ratios suggest that the
pressures in the starburst region, which are ultimately driving the
observed outflow, are very similar to those in low-redshift starbursts.

This may have a non-negligible impact on our interpretation of the
physical processes in high-redshift galaxies, because much of our
well-established low-redshift emission line diagnostics relies on
the physical gas conditions in rather subtle ways. For example, when
relating the ratios of strong, low-ionization nebular emission lines
with H$\alpha$, like, \NII/H$\alpha$, \OI/H$\alpha$, or \SII/H$\alpha$
with the \OIII/H$\beta$ ratios, starburst galaxies will fall onto a
characteristic curve \citep{BPT81,VO87}, whereas AGN will fall into a
different part of the diagram.  For high-redshift galaxies, however, the
same relationship may not be strictly valid. \citet{erb06a} observed that,
although UV selected galaxies at z$\sim 2$ follow the overall shape of the
starburst-curve in the \NII/H$\alpha$ versus \OIII/H$\beta$ diagram, as a
whole, they are offset towards larger \NII\ fluxes. \citet{brinchmann08}
collected a comparison sample of low-redshift SDSS galaxies with similar
offsets, and found that these galaxies also had statistically higher
H$\alpha$ equivalent widths. They argue that the most likely explanation
may be higher ionization parameters and densities in high-redshift
star-forming regions compared to local galaxies. We will further
develop these arguments and illustrate that pressures induced by the
starbursts may very naturally explain some of the kinematic properties
of high-redshift galaxies as well.

\begin{figure*}
\plotone{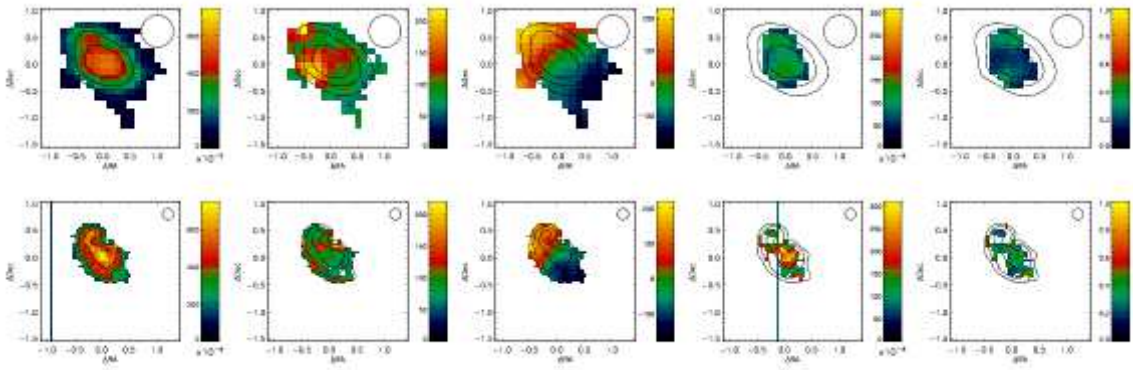}
\caption{The various emission line and kinematic maps for the source,
ZC782941, used in our analysis.  {\it{(top row and from left to right)}}
The H$\alpha$ flux, velocity dispersion, velocity, [NII]$\lambda$6583,
and [NII]$\lambda$6583/H$\alpha$ ratio maps respectively for the
250 mas (non-AO assisted) data for ZC782941. {\it{(bottom row)}} The
maps of ZC782941 in the same order as for the top row but now for the
observations taken with the 100 mas pixel scale and with the assistance
of adaptive optics. In each of the panels, we show the approximately
size of the full-width at half-maximum of the seeing disk (the disk
marked ``seeing'' in each plot) and the relative positions are in arc
seconds. The contours in each maps are the distribution of the H$\alpha$
emission in each of the two cubes (with the lowest contour and step between
contours of 1.4 $\times$ 10$^{-16}$ erg s$^{-1}$ cm$^{-2}$ arcsec$^{-2}$
for the non-AO assisted data and 1.7 $\times$ 10$^{-16}$ erg s$^{-1}$ cm$^{-2}$
arcsec$^{-2}$ for the AO assisted data). All maps were centered on the peak
in the H$\alpha$ surface brightness distribution which defines the (0,0)
in each panel.}\label{fig:ZCmaps}
\end{figure*}

However, the line ratios indicative of excitation by an AGN may
also shift at high redshift \citep{groves06}. Since most luminous
AGN at low redshift preferentially reside in massive galaxies
\citep[e.g.,][]{kauffmann03}, and because gas metallicity correlates with
galaxy mass \citep[e.g.,][]{tremonti04}, most AGN hosts at low redshift
will have narrow-line regions with high metallicities. \citet{groves06}
modeled the diagnostic line ratios for AGN in relatively low-metallicity
host galaxies.  They showed that star-formation plus AGN in relatively low
metallicity host galaxies could explain the offset observed for example
by \citet{erb06a}.  We will show that many z$\sim 2-3$ galaxies fall very
close to these regions, which may make it difficult to robustly quantify
the role of AGN and starbursts in exciting the optical emission line gas.

We will in the following present an analysis of 11 galaxies with
particularly deep near-infrared integral-field spectroscopy obtained
at the VLT. These data allow in particular to trace intrinsically
fainter nebular emission lines such as \OI$\lambda$6300 and
\SII$\lambda\lambda$6716,6731 which being close in wavelength to
H$\alpha$ often fall into the same band. A subset of 5 galaxies also have
measurements of the \OIII$\lambda$4959,5007 and H$\beta$ lines, which
fall into the near-infrared H band, and which have not been discussed
previously in the literature.  Many of the arguments in previous papers
are based on the implicit assumption that most of the emission originates
from ``ordinary'' \HII\ regions similar to low-redshift disk galaxies
and thus, that observations of the warm ionized gas are representative
of all of the kinematics and other properties of the phases of the ISM.
In contrast, our analysis is aimed at quantifying the physical conditions
in the interstellar medium of these galaxies principally by investigating
the surface brightnesses of the recombination lines, various optical
emission line ratios, and line widths to test this underlying assumption.

Throughout the paper we adopt a flat H$_0 =$70 km s$^{-1}$ Mpc$^{-3}$
concordance cosmology with $\Omega_{\Lambda} = 0.7$ and $\Omega_{M} =
0.3$. 

\section{Observations and data reduction}
\label{sec:observations}

For our analysis of the physical conditions in the interstellar
medium of strongly star-forming, high-redshift galaxies, we collected
a sample of 11 galaxies at z$\sim$1.5$-$2.5 from the SINS program
\citep{fs06} with rest-frame optical integral-field spectroscopy
(Table~\ref{table:properties}).

Data were taken with the near-infrared integral-field spectrograph
SINFONI on the ESO Very Large Telescope in several runs between 2004
and 2006. Observations have been presented elsewhere \citep[][]{fs06,
genzel08}. We reduced these data independently from any other previous
work, using the IRAF \citep{tody93} standard tools for the reduction
of longslit spectra, modified to meet the special requirements of
integral-field spectroscopy, and complemented by a dedicated set of
IDL routines. Data are dark-frame subtracted and flat-fielded. The
position of each slitlet is measured from a set of standard SINFONI
calibration data, measuring the position of an artificial point
source. Rectification along the spectral dimension and wavelength
calibration are done before night sky subtraction to account for some
spectral flexure between the frames. Curvature is measured and removed
using an arc lamp, before shifting the spectra to an absolute (vacuum)
wavelength scale with reference to the OH lines in the data. To account
for variations in the night sky emission, we normalize the sky frame to
the average of the object frame separately for each wavelength before
sky subtraction, correcting for residuals of the background subtraction
and uncertainties in the flux calibration by subsequently subtracting the
(empty sky) background separately from each wavelength plane.  These data
reduction procedures have fewer interpolations and are optimized for
faint, extended, low surface brightness objects compared to the SINFONI
pipeline which obviously must be capable of reducing a much wider variety
of objects.  Overall, we expect that these differences lead to more
robust data compared to previous reductions but overall the results are
largely consistent.

The three-dimensional data are then reconstructed and spatially aligned
using the telescope offsets as recorded in the header within the same
sequence of dithered exposures (about one hour of exposure), and by
cross-correlating the line images from the combined data in each sequence,
to eliminate relative offsets between different sequences. Telluric
correction is applied to each final cube. Flux scales are obtained from
standard star observations taken every hour at similar position and air
mass as the source.

\begin{figure*}
\plotone{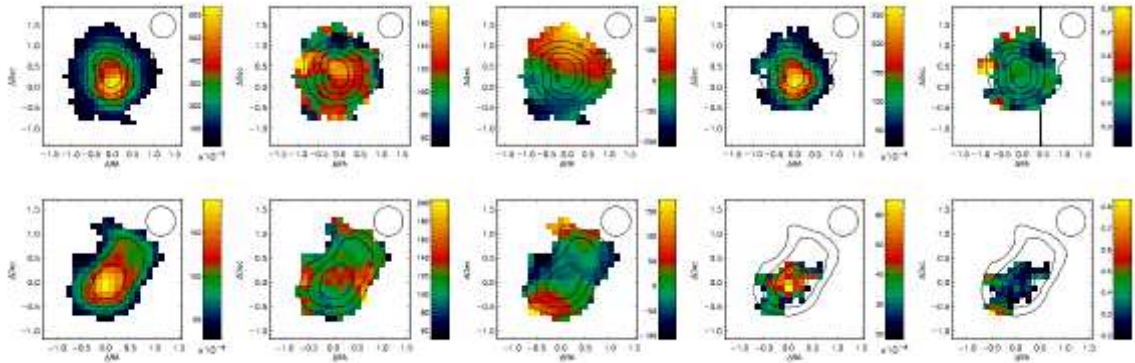}
\caption{The various emission line and kinematic maps for the source,
Q2343-BX610 and Q2343-BX528, used in our analysis.  We show these two
sources for comparison with previously published maps in the literature.
{\it{(top row and from left to right)}} The H$\alpha$ flux, velocity
dispersion, velocity, [NII]$\lambda$6583, and [NII]$\lambda$6583/H$\alpha$
ratio maps respectively for the for Q2343-BX610. {\it{(bottom row)}}
The maps of Q2343-BX528 in the same order as for the top row.  In each
of the panels, we show the approximately size of the full-width at
half-maximum of the seeing disk (the disk marked ``seeing'' in each
plot) and the relative positions are in arc seconds. The contours
for the maps of Q2343-BX610 represent the continuum, while those of
Q2343-BX528 represent the H$\alpha$ surface brightness (with the lowest
contour and step between contours of 4 $\times$ 10$^{-17}$ erg s$^{-1}$
cm$^{-2}$ arcsec$^{-2}$). All maps were centered on the peak in the
H$\alpha$ surface brightness distribution which defines the (0,0) in
each panel.}\label{fig:BX610-528maps} \end{figure*}

\begin{figure*}
\plottwo{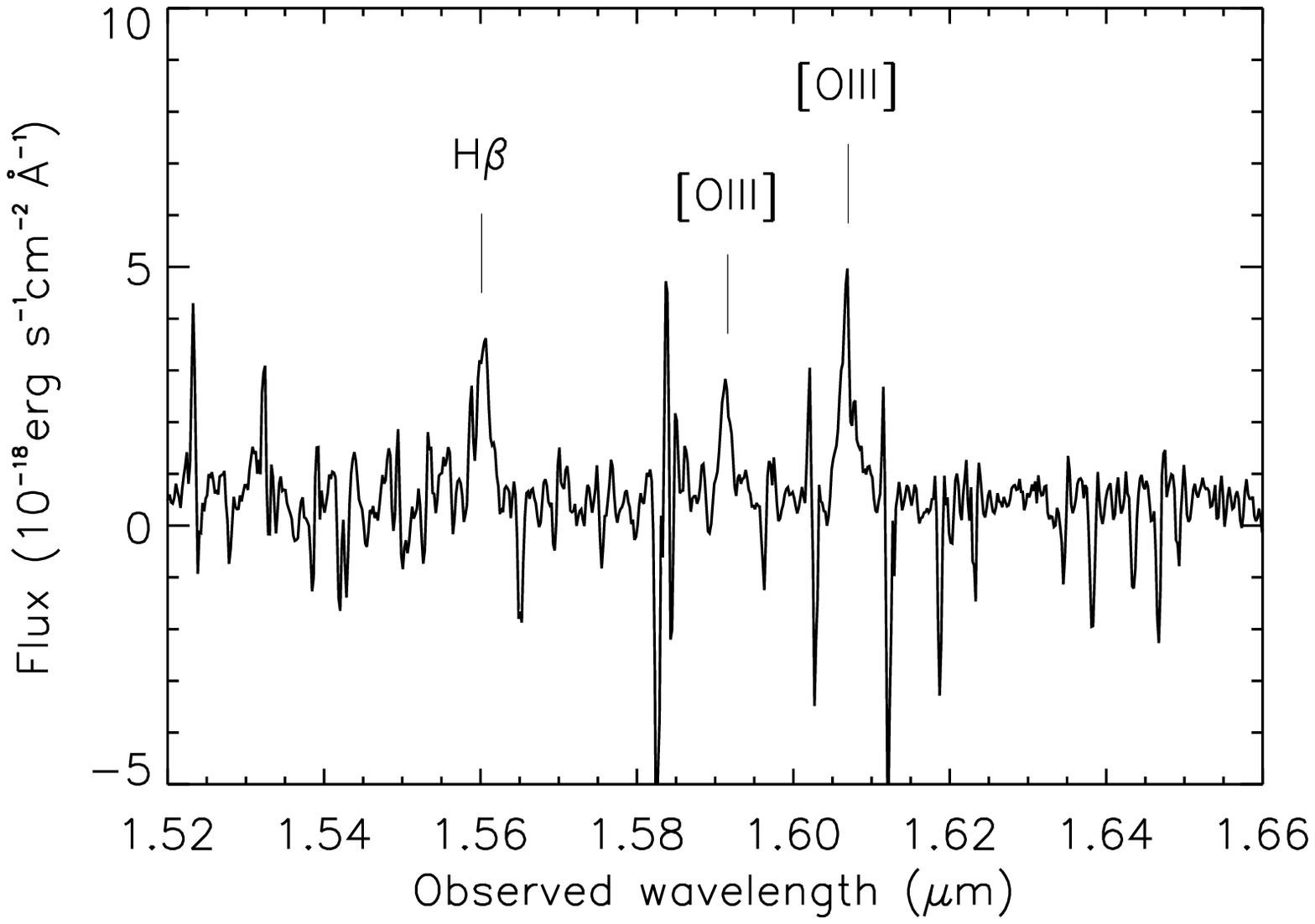}{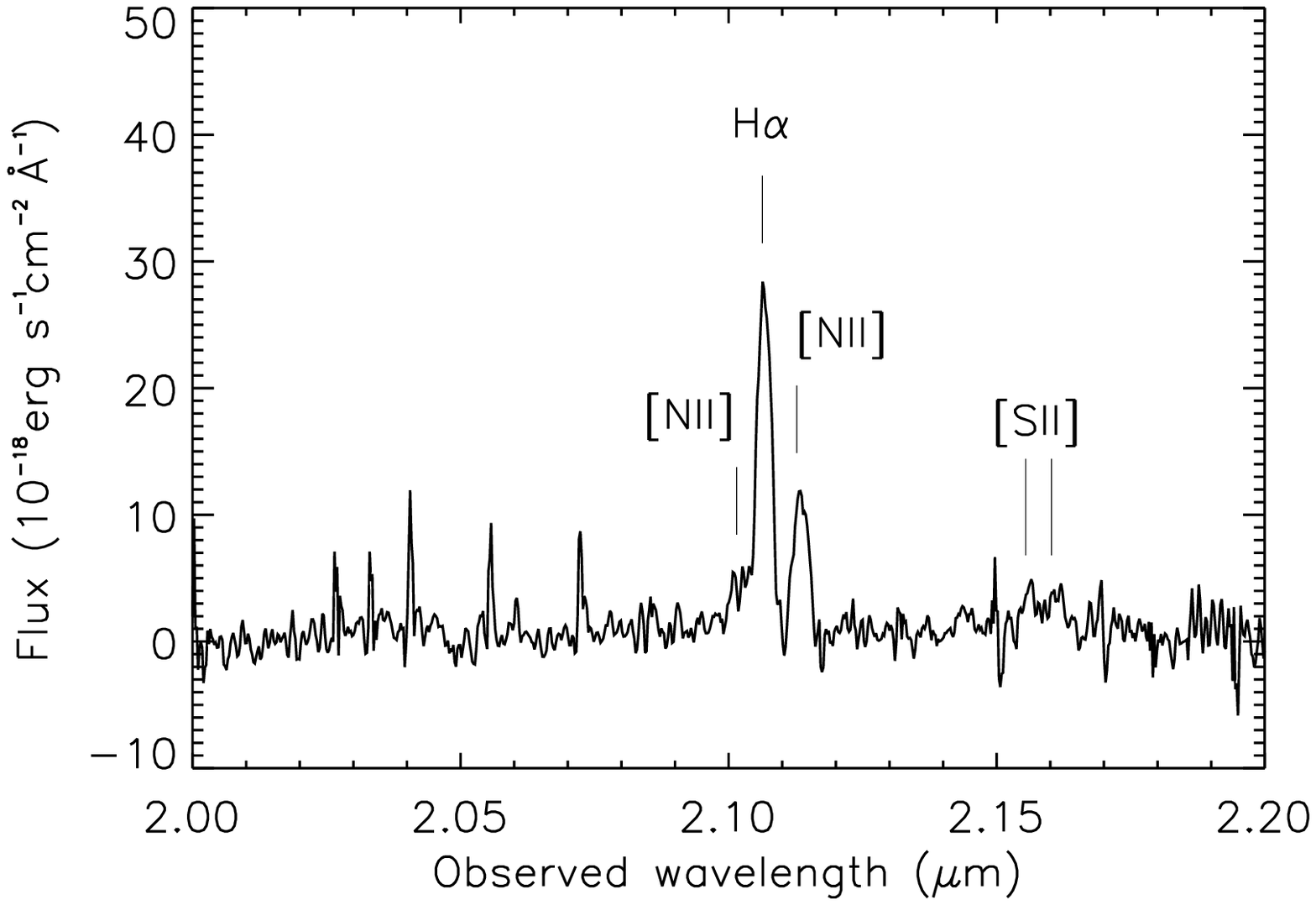}
\caption{The integrated spectrum of Q2343-BX610 in the H-band
{\it{(left)}} and the K-band {\it{(right)}}. All of the strong optical
lines are indicated.}\label{610intspec}
\end{figure*}

We also used the standard stars to monitor the seeing during observations,
and we find an effective seeing in the combined cubes of typically FWHM
\as{0}{5}$-$\as{0}{8}. The spectral resolution was measured from night-sky
lines and is FWHM$\sim$ 115 and 150 km s$^{-1}$ in the K and H-bands,
respectively, for the 250 mas pixel scale (``mas'' is milli-arcseconds).
In addition, for two of the sources, data were taken with adaptive optics
assistance which yielded a seeing about 200 mas.  The data were taken
with the 100 mas pixel scale in SINFONI.  In Fig.~\ref{fig:ZCmaps} we show
an example of various maps that have not been presented previously for
the source, ZC782941.  In addition, we show example maps of the sources,
Q2343-BX610 and Q2343-BX528 for comparison with previously published
maps \citep[e.g.,][]{fs06}.  In our subsequent analysis, we will analyze
the integrated H-band spectra of the sources when they are available.
We show an example of the H and K-band integrated spectra for the source,
Q2343-BX610, in Fig.~\ref{610intspec}.

In addition, we reduced publicly available HST/NICMOS images of 5 of
the galaxies in our sample, which were observed as part of proposal
ID 10924 (P.I. Shapley). Each galaxy was observed for 4 orbits using
the NIC2 camera with the F160W (H-band) filter and a pixel scale of
\as{0}{075}. The calibrated individual exposures (\_cal files) were
downloaded from the HST archive and reduced using the standard IRAF
routines {\it{pedsky}}, to correct the well-studied pedestal effect
caused by residual bias, and {\it{multidrizzle}}, to combine the
exposures. Images were corrected for the impact of additional cosmic
ray events during passage through the South Atlantic Anomaly when
required. The resulting images are shown in Fig.~\ref{fig:nicmosimages}.

\section{The remarkably high H$\alpha$ surface brightnesses and trends}
\label{sec:results}

In this section, we will discuss the emission line properties of
our sample of galaxies such as surface brightnesses, the relationship
between the velocity dispersion and surface brightness, and the impact of
``beam smearing''.  This analysis will form the basis of our subsequent
discussion. We refer the reader to \citet{fs06}, \citet{genzel06}, and
\citet{genzel08} for a discussion of the overall kinematic properties
of the galaxies in our sample.

\subsection{H$\alpha$ surface brightness}\label{subsec:HaSB}

One of the most remarkable, but so far little discussed, observational
findings of z$\sim 2-3$ galaxies studied with IFUs are their
high emission line surface brightnesses. In our sample we measure
H$\alpha$ surface brightnesses of $\sim$10$^{-15.1}-10^{-16.9}$ erg
cm$^{-2}$ s$^{-1}$ arcsec$^{-2}$ (Fig.~\ref{fig:SBHaradius}). The
data were smoothed by 3$\times$3 pixels, i.e., averaged in areas
of size \as{0}{375}$\times$\as{0}{375}, which is appropriate since
the spatial resolution is typically $\sim$ \as{0}{5}-\as{0}{6} or
$\sim$4$-$5 pixels. Correcting for cosmological surface brightness
dimming, this corresponds to a typical rest-frame surface brightness
of 10$^{-13.2} - 10^{-14.8}$ erg cm$^{-2}$ s$^{-1}$$\sq \arcsec$
(Fig.~\ref{fig:HaSBsigma}). This represents a lower limit to the highest
intensity regions, due to the ``beam smearing'' effect of the low spatial
resolution of our data. In fact, for two of the galaxies, we can further
quantify the effect of beam smearing on both the surface brightness
distribution and line widths. For BzK-15504 and ZC782941, where we have
both adaptive optics assisted and seeing limited observations, we find
that the offset in surface brightness between both sets of observations
is typically more than a factor of 3 (Fig.~\ref{fig:AOHaSBsigma}). This
factor is similar to the ratio of the area of the point spread function in
both sets (the full width half maximum of the seeing disk is typically
about \as{0}{5} for the seeing limited data versus about \as{0}{2} in
the AO assisted data), suggesting that the star-forming clumps are at
best marginally resolved in either data set. Much of the structure must
be smaller than about 2 kpc, the physical resolution of our highest
resolution data. Moreover, this implies that all of the measured
surface brightnesses of the most intense H$\alpha$ emission regions are
under-estimated, even in the AO-assisted observations.

\begin{figure*}
\plotone{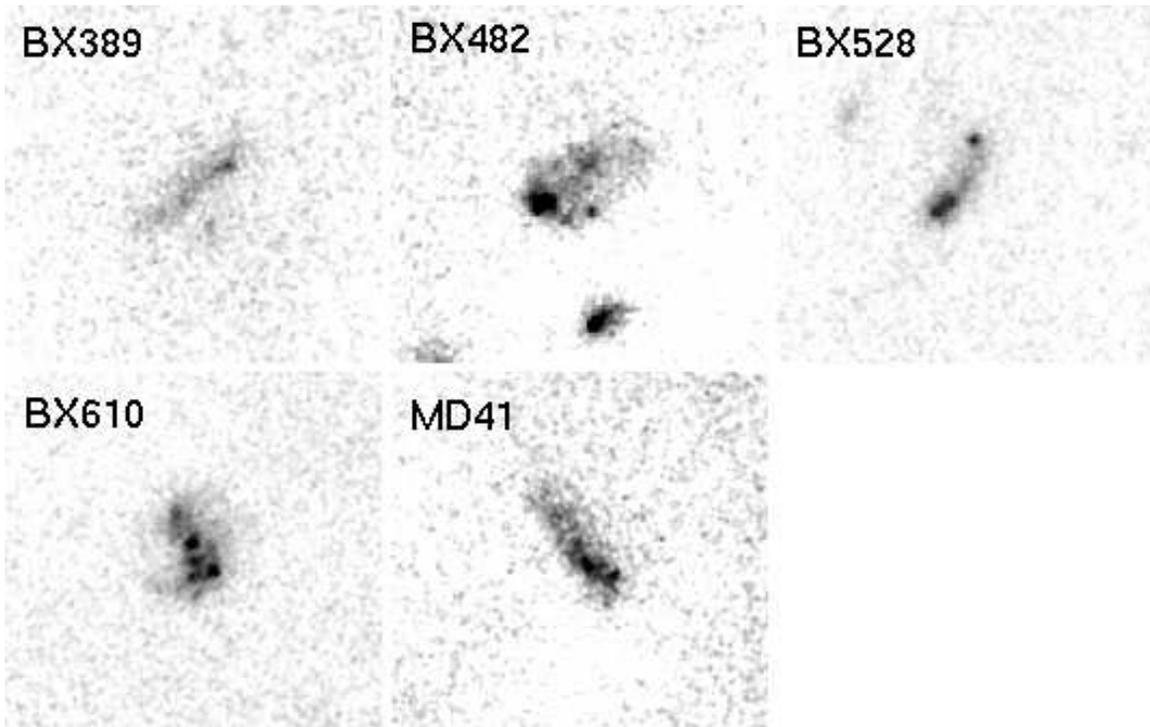}
\caption{HST/NICMOS NIC2 H-band (F160W) images of 5 galaxies within
the sample. Images are 5$\times$5$''$, and North is at the top and East
is on the left.  The images have a pixel scale of 30 mas (the drizzled
pixel size) and each image reaches a depth of about 28.6-28.7 m$_{AB}$
pixel$^{-1}$ (for a NICMOS pixel scale of 76 mas) or about 25.8-25.9
m$_{AB}$ arcsec$^{-2}$.}
\label{fig:nicmosimages}
\end{figure*}

An H$\alpha$ surface brightness comparison between galaxies in the local
and distant Universe illustrates the exceptional nature of the galaxies
in our sample. Galaxies that reach the highest surface brightnesses we
observe and over similar physical scales do not exist at low redshift.
In a study of 84 Virgo cluster and isolated spiral galaxies, many
galaxies reach surface brightness levels at the low end of what we
have observed at high redshift ($\log \rm SB_{H\alpha} \lesssim -14$
erg cm$^{-2}$ arcsec$^{-2}$) but only in their nuclei and on
scales about or less than 1~kpc \citep[][]{koopmann06a, koopmann06b,
koopmann04a, koopmann04b, koopmann01}. Local starburst galaxies can
reach higher surface brightnesses \citep[but not as high as our peak
surface brightnesses,][]{lehnert95, lehnert96a} but again only in
regions that are nuclear or circum-nuclear with sizes $\lesssim$ 1
kpc \citep[and well within the ``turn-over'' radius of the rotation
curve;][]{lehnert96b}. Even in the more extreme starburst sample of
\citet{armus89,armus90}, only a handful of galaxies reach surface
brightnesses sufficient to be observed at z$\sim$2 (such as M82). At
high redshift, H$\alpha$ surface brightnesses as extreme as those found
only in the nuclei of nearby galaxies on small scales are found over
significantly larger isophotal radii of order 10-20 kpc and are generally
more extreme. Although our comparison includes some of the most powerful
and intense starbursts in the local Universe, they generally do not
reach sufficiently high surface brightnesses over large enough areas to
correspond to any of the galaxies in our distant galaxy sample. Moreover,
at the 4$-$5 kpc resolution of our seeing limited high-redshift data,
none of the local starbursts would reach these maxima in the surface
brightnesses that are observed due to the heavy spatial smoothing and
dilution by regions with lower surface brightness.

This lack of correspondence renders any simple analogy between
``ordinary'' quiescently star-forming spiral galaxies in the local
Universe and galaxies at z$\sim$2 questionable. Possible significant
differences include the distribution of mass in various phases of the
ISM, of density, of star-formation intensity, of gas fraction, and many
more. Moreover, this lack implies that we cannot only study the kinematics
of z$\sim$2 galaxies without investigating the nature of the physical
processes that power their high surface brightness line emission, and
the impact such a finding has on their kinematics and overall gas physics.

\subsection{ H$\alpha$ surface brightness: Characteristic correlations
and Beam Smearing} \label{SBcharacteristics}

\begin{figure}
\plotone{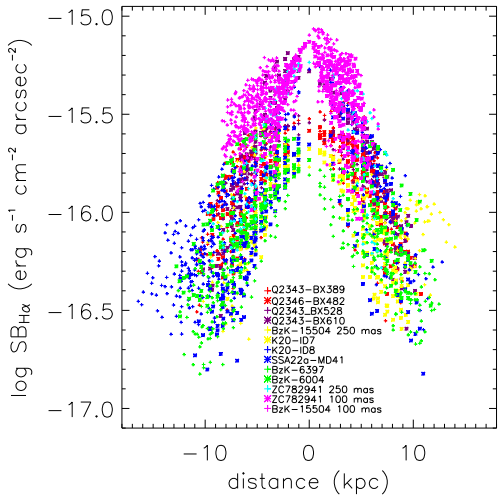}
\caption{A plot of the observed H$\alpha$ surface brightness versus
the projected physical radius. The surface brightness has not been
corrected for cosmological dimming.  Each point represents a position
centered on one pixel of \as{0}{125}$\times$\as{0}{125} , whereas the
data were smoothed by 3$\times$3 pixels, i.e., averaged in areas of size
\as{0}{375}$\times$\as{0}{375}. As the spatial resolution is typically
$\sim$ \as{0}{5}-\as{0}{6} (or $\sim$4$-$5 pixels), the points are
therefore not independent, and each data set has a limited number of
independent points \citep[about 5-10, see ][]{shapiro08}. The legend to
the figure indicates each galaxy in our sample.  The zero radius was
chosen to represent the symmetry point in either the kinematics or the
outer isophotal H$\alpha$ distribution. The distribution of surface
brightness is not symmetrical for most of the sources.}
\label{fig:SBHaradius}
\end{figure}

\begin{figure}
\plotone{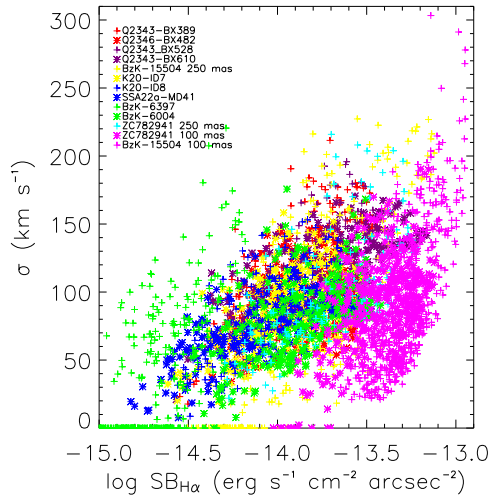}
\caption{Plot of the H$\alpha$ surface brightness (now corrected for
surface brightness dimming) versus the observed velocity dispersion. See
Fig.~\ref{fig:SBHaradius} for details on the 3$\times$3 pixel
smoothing that was applied. The legend to the figure indicates which
symbol represents which galaxy in our sample.  There appears to be a
relationship between surface brightness in H$\alpha$ and the line width
for both the ensemble of sources as well as within individual sources.
Because of the trend between surface brightness and dispersion, velocity
dispersions below the resolution of our data, about 50 km s$^{-1}$ for
the sources observed in the K-band (all but one of the sources shown
here) and about 60 km s$^{-1}$ for BzK-6397 have particularly large
relative uncertainties.}
\label{fig:HaSBsigma}
\end{figure}

\begin{figure}
\plotone{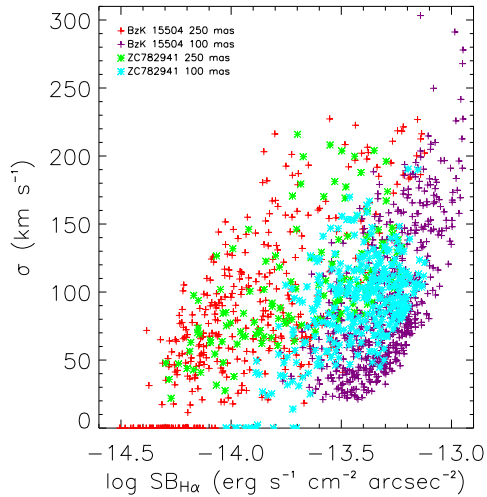}
\caption{Plot of the rest-frame H$\alpha$ surface brightness versus
the observed velocity dispersion for the 2 galaxies for which we have
both seeing limited and adaptive optics assisted data sets (ZC782941 and
BzK-15504).  Individual points in the data cubes were treated as shown in
Fig.~\ref{fig:HaSBsigma}.  In the case of the data cubes that were taken
with adaptive optics assistance, the pixel scale is 50 mas pixel$^{-1}$
and they were also averaged over 3$\times$3 pixels (Fig.~\ref{fig:S
BHaradius}.  There is an increase in the velocity dispersion of some of
the regions within BzK-15504 but these are entirely associated with the
region of and around the AGN, substantiating our claim that BzK-15504
hosts an AGN \citep[see also][]{genzel06}.}
\label{fig:AOHaSBsigma}
\end{figure}

To elucidate the underlying physical cause of the high H$\alpha$ surface
brightness of our galaxies, we searched for correlations with other
parameters. As shown in Fig.~\ref{fig:SBHaradius}, \ref{fig:HaSBsigma},
and \ref{fig:AOHaSBsigma}, we find that both the H$\alpha$ surface brightness
and H$\alpha$ velocity dispersion decline with radius. We
generally do not find a substantial increase in line widths when comparing
the seeing-limited data with data taken with the adaptive optics system,
although as expected, we observe higher H$\alpha$ surface brightness in
some regions (Fig.~\ref{fig:AOHaSBsigma}). Thus beam smearing does have
a tendency to lower the observed surface brightness suggesting that most
of the structure within these galaxies is not resolved.

However, beam smearing is worrying in that the trends we observe will
obviously be influenced by our poor resolution. We must take care in
determining what the true impact of our poor resolution might actually be
on the distribution and kinematics of the emission line gas. We observe
the trends with H$\alpha$ surface brightness over physical scales that
are $\sim$6$\times$ larger than the spatial resolution of our data. To
investigate explicitly whether this may be an artifact due to low spatial
resolution we constructed several simple models that have bright central
point sources with broad lines. Such models would correspond to a bright
AGN or concentration of mass increasing the velocity dispersion in the
centers of these galaxies.  Although this may appear unphysical, since
we do not see bright point sources, it will help to investigate, for
example, the effects of a narrow-line AGN. Our toy models (not shown) do
not reproduce the observed trends, which implies it is very difficult to
contrive a realistic situation where only the poor resolution of our data
would lead to the trends we observe between surface brightness and radius.

Could beam smearing induce an apparent relationship between line width
and surface brightness? Could this trend be related to distant galaxies
having more complex light profiles? To understand the impact of beam
smearing and complex light profiles of these galaxies on our analysis,
we constructed simulated data cubes using the light distributions
observed in NICMOS H-band images of five galaxies in our sample
(Fig.~\ref{fig:nicmosimages}).  We assumed that the line emission follows
the distribution of the H-band flux, that the velocity dispersion is 25
km s$^{-1}$ independent of position or radius \citep{epinat09}, and that
the rotation curves and peak velocities are those from \citet{fs06}.
We did not include noise in this display as it results in a scatter
plot about the assumed constant velocity dispersion, masking the trend
of some pixels to reach high dispersions (which we think is extremely
important to make obvious). Thus for clarity, we do not show a plot of
our analysis with noise.

\begin{figure}
\plotone{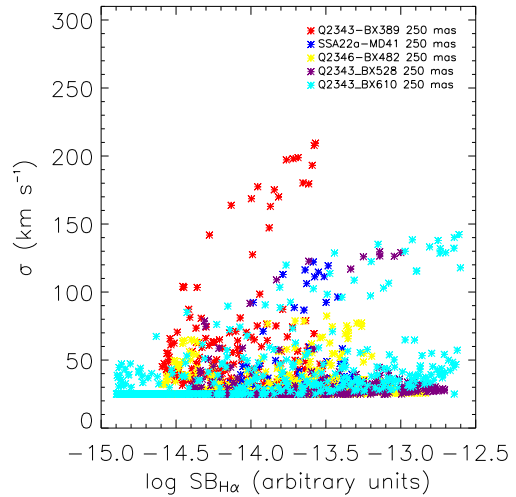}
\caption{A plot of surface brightness versus velocity dispersion for a
set of toy models.  Each of the ``models'' was generated assuming the
distribution of light of the HST/NICMOS image, an intrinsic and uniform
velocity dispersion of 25 km s$^{-1}$, a distribution of rotation
velocity and a velocity peak as given in \citet{fs06}, and a point
spread function with a FWHM of \as{0}{6}. Care was taken to ensure
that the final resolution of the data set was \as{0}{6} and took into
account the intrinsic resolution of NICMOS. For the presentation here,
we did not include noise as it would simply tend to increase the scatter,
especially at the low surface brightnesses, but not change the overall
trend that the dispersion would remain roughly constant.}
\label{fig:toyHaSBsigma}
\end{figure}

As we can see in Fig.~\ref{fig:toyHaSBsigma}, assuming typical velocity
dispersions seen in local disk galaxies and light distributions of distant
galaxies and smoothing them to our resolutions does not reproduce the
data \citep[see also][]{wright08, genzel08, fs06}.  We do see that
there are some regions of high dispersion, but this makes up a small
number of pixels whereas a great majority of the regions, regardless of
their relative surface brightness have dispersions similar to what we
initially assumed (i.e., 25 km s$^{-1}$). This, coupled with the lack of
increase in the dispersions when comparing our seeing limited data with
that taken using adaptive optics, suggests that beam smearing, while
obviously playing a role in these trends, does not cause these trends.

\section{Emission-line properties of high-redshift galaxies}

Given the high star-formation rates \citep{fs06, genzel08} and emission
line surface brightnesses in these galaxies, we hypothesize that the
high surface brightnesses and relationship between velocity dispersion
and surface brightness in H$\alpha$, may be explained by postulating
that the intense star formation is pressure-driven by mechanical energy
input from the starburst itself and self-gravity of gas at high surface
densities. In this sense, the star formation will be self-regulated. We
will argue that these systems may be analogous to nearby starburst
galaxies, such as  M82, except that at high redshift, the intense
star formation and strong mechanical energy injection must act over
significantly larger areas \citep[10-20 kpc compared to a few kpc,
e.g.,][]{heckman90, lehnert96a}, but with a similar local surface
brightness and similarly high pressures. However, this is not the
only possibility and we will explicitly address different scenarios to
investigate whether these relationship may be generated by cosmological
gas accretion or gravitationally unstable disks.  Some of our main
arguments will rely on the detection of nebular emission lines like
[NII]$\lambda$6583, [SII]$\lambda$$\lambda$6716,6731 or [OI]$\lambda$6300,
which are relatively faint in high-redshift galaxies. We detected all of
these lines in only one galaxy, while we detect all but [OI]$\lambda$6300
in 5 others. Where one of the lines is undetected or severely affected
by a telluric night sky line, we give upper limits, provided that they
are physically meaningful.  For a subsample of 5 targets, our data
include H$\beta$ and [OIII]$\lambda\lambda$4959,5007, which fall into
the near-infrared H-band for redshifts z$>$2. These data sets have not
been discussed previously in the literature.  We list the integrated
emission line properties of these galaxies in Table 1. As the H-band data
are rather shallow relative to the K-band, we typically detect
only the highest surface brightness regions (and likely also regions of
relatively low extinction).

We extracted emission line ratios from matched apertures in the H and
K band data sets, covering the areas where line emission is detected
in the H band. We use the measured, and not the extinction corrected
line fluxes for the emission line diagnostics. This adds only a minor
uncertainty, given the relatively low signal-to-noise ratio of our data,
low luminosities of the low-ionization lines, and the fact that we will
only use ratios of lines with very similar rest-frame wavelengths.

\subsection{Extinctions}
\label{ssec:extinction}

We estimated extinctions for the four objects with measured H$\alpha$
and H$\beta$ fluxes (Table 2). H$\beta$ in Q2346-BX482, which would have
been a fifth galaxy with a H$\beta$ flux and extinction estimate, and the
emission in the north of Q2343-BX389 are severely affected by night sky
lines (so the measurement is only for the southern part of the galaxy),
and for three other galaxies we identified the areas with reasonably
bright H$\beta$ emission, and extracted H$\alpha$ from the same aperture
(Table 2).  For a galactic extinction law, and an intrinsic Balmer ratio
of F$_{H\alpha}$ /F$_{H\beta}$ $=$2.86, we find extinctions in the range
of A$_{V}\sim$ 1$-$2 magnitudes. This corresponds to correction factors of
about 2$-$5 between observed and intrinsic H$\alpha$ luminosities. We will
neglect extinction in much of our subsequent discussion, but caution that
intrinsic values are strict lower limits and may have been underestimated
by factors of a few.

\subsection{Electron densities and pressures} 
\label{ssec:density}

The line ratio of the \SII\ doublet, \SII$\lambda$6716/\SII$\lambda$6731,
is density-sensitive in the range of $\sim 10^{1-5}$ cm$^{-3}$. These
lines are relatively faint, but we have robustly detected and spectrally
resolved this doublet in six galaxies in our sample, at signal-to-noise
ratios $\sim10$. We are thus able to measure electron densities directly,
and to estimate the pressure in the partially ionized zones within the
interstellar medium. Densities listed in Table~\ref{table:SII} are in the
range $\sim$ 100$-$1000 cm$^{-3}$. Such values are typical, if not higher,
than densities in the starburst regions of galaxies in the local Universe
\citep[such as M82; ][]{lehnert96a} and suggest thermal pressures of about
10$^{-8.5}-10^{-9.5}$ dyne cm$^{-2}$, i.e., several orders of magnitude
higher than in the interstellar medium in normal nearby galaxies. Similar
pressures are found for the z$=$2.6 submillimeter galaxy SMMJ14011+0252
\citep{nesvadba07}. We will argue below that the ambient medium of our
galaxies is highly pressurized, and show that these estimates are also
consistent with pressures derived directly from photoionization models.

The high pressures may also explain another remarkable feature of our
galaxies, namely their overall low ratios of low-ionization lines like
\SII$\lambda\lambda$6716,6731 and \NII$\lambda$6583 compared to the
high H$\alpha$ luminosities. While the recombination lines increase
linearly with increasing density and ionization parameter, the ratio of
\SII$\lambda\lambda$6716,6731 will decline with increasing ionization
parameter \citep[e.g.,][]{wang99}. Since the gas with the highest
surface brightness also has a declining ratio, measuring the density
from the \SII\ lines becomes more difficult as the surface brightness
increases. Thus, it may not be surprising that we obtained \SII\
measurements at sufficiently high signal-to-noise only for parts of our
sample. This may also affect other diagnostic line ratios. Larger samples
of high-redshift galaxies with deep spectroscopy of a comprehensive set
of faint, diagnostic lines will help to secure these findings.

\subsection{Diagnostic line ratios}

\citet{BPT81} and \citet{VO87} advocated the use of characteristic
ratios of the bright optical nebular emission lines as diagnostics
to differentiate between ionization due to starbursts and
AGN. These so-called BPT diagrams relate the strengths of lines like
\NII$\lambda$6583, \OI$\lambda$6300, or \SII$\lambda\lambda$6716,6731
with those of the Balmer recombination lines and \OIII$\lambda$5007,
and gives us the ability to trace the physical conditions, namely
temperature and ionization parameter, in the emission line gas. Due to
the different ionizing spectra of starbursts and AGN, galaxies will fall
into characteristic areas of the diagrams when their nebular emission
is dominated by photoionization from young stars, or by an active nucleus.

\begin{figure}
\plotone{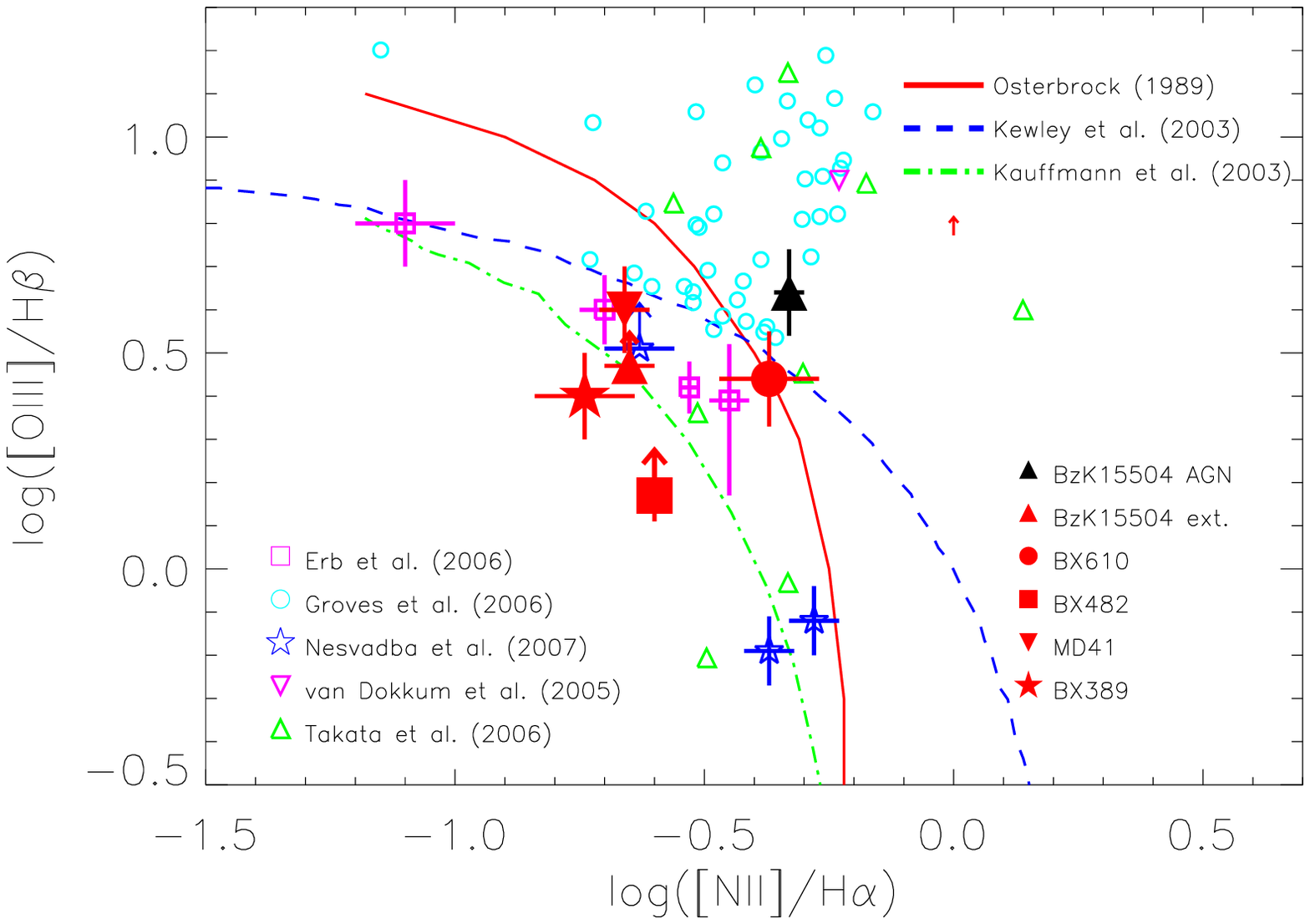}
\caption{The optical emission line ratios, \NII$\lambda$6583/H$\alpha$
versus \OIII$\lambda$5007/H$\beta$ for some of the galaxies in our sample
(BzK-15504, BX610, BX482, MD41, and BX389), as well as for some distant
and local galaxies taken from the literature \citep{erb06a, groves06,
nesvadba07, vandokkum05, takata06}.  The dividing lines within the
diagram demarcate the separate regions occupied by galaxies whose
emission lines are dominated by heating due to star formation from
those dominated by active galactic nuclei \citep{kauffmann03, kewley01,
osterbrock89}. Interestingly, many of the galaxies in this sample fall
near the dividing line, which  is consistent with either heating due to
AGN, as is the case for both the nucleus and the extended emission line
gas in BzK-15506.}
\label{fig:BPTNIIHa}
\end{figure}

We show the results in Fig.~\ref{fig:BPTNIIHa} and \ref{fig:BPTSIIHaOIHa},
together with the loci of other galaxies at similarly high redshifts taken
from the literature. It should be noted that all of these diagnostics are
to a large degree empirical and developed for galaxies at low redshift,
and that different evolutionary stages may influence the emission line
diagnostics in a rather subtle way. As already stated by \citet{erb06a},
many z$\sim 2-3$ galaxies are shifted towards higher low-ionization line
ratios relative to the low-redshift relationships. \citet{brinchmann08}
argued by analogy with a subsample of local galaxies from the SDSS
that this may be a result of higher pressures and higher ionization in
high redshift galaxies.

However, this is not the only effect we may expect to take place. At low
redshift, luminous AGNs reside predominantly in relatively massive galaxies
which have comparably high ($\ga$solar) metallicities. At high redshift,
this is not necessarily the case. \citet{groves06} modeled the expected
position of AGN in low-metallicity host galaxies in classical emission
line/ionization diagrams (e.g., \NII/H$\alpha$ versus [OIII]/H$\beta$).
They find that the metallicity-sensitive \NII/H$\alpha$ ratio will be
shifted towards lower values for AGN with low-metallicity narrow line
regions. Seyfert galaxies and QSOs have ratios of [OIII]/H$\beta$ that
are higher than those of HII regions and star-forming galaxies in such
diagrams. Since the ratio of, for example, [OIII]/H$\beta$ is not strongly
affected by lower the metallicity, the position of low-metallicity
AGN in ionization diagrams may lie above the locus occupied by
HII regions.  We illustrate this effect by showing the position of
the low-redshift, low-metallicity AGN of \citet{groves06} in the same
diagram. Interestingly, most optically or UV-selected starbursts fall very
close to the region spanned by the \citeauthor{groves06} sample. This
is not to say that these will be low-metallicity AGN, in particular,
since the observed strong starbursts will create high-ionization,
high-pressure environments, which will shift the galaxies towards the
AGN part of the diagram \citep{brinchmann08}. However, this does imply
that we may significantly underestimate the fraction of high-redshift
galaxies with AGN of rather moderate luminosity. In the following we
discuss the specific example of an obscured quasar with relatively
inauspicious diagnostic line ratios compared to optically selected QSOs.

\section{The powerful Active Galactic Nucleus of BzK-15504}
\label{sec:brazil}

We will now give an example from among the galaxies of our sample,
BzK-15504, where the AGN plays a role that is almost certainly
non-negligible in interpreting the extended emission of the host
galaxy. This galaxy falls within the range of low-metallicity AGN in
the \NII/H$\alpha$ versus \OIII$/$H$\beta$ diagnostic diagram discussed
by \citet{groves06}, but also near the dividing line between AGN and
starbursts. Thus, the impact of AGN photoionization will easily be missed
with a diagnostic based only on local galaxies since powerful AGN embedded
in low metallicity host galaxies are relatively rare \citep{groves06}.
We use additional constraints, in particular the \OI/H$\alpha$ ratio
and the near-nuclear \OIII$\lambda$5007 luminosity to argue that this
galaxy hosts an obscured quasar, and that most of the extended H$\alpha$
emission may in fact be part of an AGN ionization cone. This calls into
question previous interpretations of the star-formation properties and
accretion rates in this galaxy (see also \S~\ref{ssec:implications}). While
this is the only clear case among the 11 galaxies discussed here, it
does illustrate that the rest-frame optical emission line diagnostics
for z$\sim2$ galaxies, including the 11 discussed here, may be less
straight-forward to interpret than previously realized.

\subsection{The Bolometric Luminosity of the QSO in BzK-15504}

It is clear that the nuclear emission lines of BzK-15504 lie
within the region of the AGN in all of the emission line diagrams
suggesting that it is indeed a powerful AGN (Fig.~\ref{fig:BPTNIIHa}
and \ref{fig:BPTSIIHaOIHa}).  The bolometric luminosity is an important
parameter as it sets the total energy output of the AGN. BzK-15504
has not been observed over a sufficiently wide range of wavelengths to
estimate its bolometric luminosity accurately. \cite{heckman04} argue
that \OIII$\lambda$5007 luminosity can be used to estimate the bolometric
luminosity of AGN over the range of L$_{[OIII]}$=10$^{6.5}$ to 10$^9$
L$_{\sun}$. Their adopted relationship is L$_{bol}$=3500L$_{[OIII]}$
with a scatter of about a factor of 2 \citep[see][and references therein
for details]{heckman04}. Using this relation for our \OIII$\lambda$5007
flux from BzK-15504 centered on the brightest continuum and line emitting
region suggests that it has L$_{[OIII]}$=10$^{9.5}$ L$_{\sun}$ and an
implied L$_{bol}$=10$^{13.0}$ L$_{\sun}$. Extinction correcting the
\OIII$\lambda$5007 flux would increase it by 0.5 dex. BzK-15504 hosts
a powerful AGN indeed -- a QSO!

\subsection{BzK-15504 as a giant Narrow Line Region?}

In addition to the nuclear line ratios, the line ratios of the extended
emission in BzK-15504 suggest that it could be photoionized by the
AGN. In Fig.~\ref{fig:BPTNIIHa} and \ref{fig:BPTSIIHaOIHa}, we show the
BPT diagrams for various line emitting regions within BzK-15504, and the
line ratios lie either near the star formation-AGN boundary, as in the
diagrams of \OIII/H$\beta$ versus \SII/H$\alpha$ and \OIII/H$\beta$ versus
\NII/H$\alpha$, or clearly within the AGN region as in the \OIII/H$\beta$
versus \OI/H$\alpha$ diagram.

This of course is perhaps obvious from the fact that the surface
brightness of the extended emission line region of BzK-15504 is very high,
one of the highest in the sample (Fig.~\ref{fig:SBHaradius}). However,
can illumination by a central AGN explain the light profile of BzK-15504?
The light profile of BzK-15504 has a ${{1}\over{r^2}}$ in its H$\alpha$
surface brightness dependence and is thus consistent with photoionization
from a point source such as an AGN. However, since the emission is
somewhat complex, this is not a strong constraint as other light profiles
might equally well fit the data. Our point here is to suggest that it
is at least ``consistent'' with a ${{1}\over{r^2}}$ profile. Making this
assumption allows us to make a rough estimate of the ionization parameter,

\begin{equation}
\bar{\rm U}_0=S_{QSO}/c \bar{\rm n}_H;\ \bar{\rm n}_H={{P_{gas}}\over{k 10^4}}
\end{equation}

where c is the speed of light, $S_{QSO}$ is the photon intensity at radius
r from the QSO, and where we used, $L_{QSO}=10^{45.5}$ erg s$^{-1}$
(or 1/10 of the total bolometric luminosity), a radius of about 6
kpc, a rough density of 500 cm$^{-3}$ and a factor of 3 between the
electron density and total density to reflect the fact that most of the
\SII$\lambda\lambda$6716,6731 forms in the partially ionized zone. Using
these numbers suggests that $\bar{\rm U}_0$ is about 10$^{-3}$. As we
will see later, this is a rather canonical number for some of the other
galaxies in the sample and not surprising given the surface brightness
of all the galaxies.

Looking at this from another perspective, we can estimate that it takes
a few percent of the total bolometric luminosity to explain the total
H$\alpha$ emission in BzK-15504, if we make a simple recombination
estimate for the number of ionizing photons. This is less than the 10\%
we assumed above, but overall consistent with the nebula being completely
powered by the QSO. We note that this is obviously a lower limit as we
have not considered the extinction in the extended nebular emission.

The total ionized gas mass is straight-forward to estimate as well. If we
assume simple case B recombination as we did previously to estimate the
total ionizing luminosity necessary to power the nebula, we find that
we need about,

\begin{eqnarray}
\nonumber
{\rm M}_{H} = {{{L_{H\alpha}}} \over {h \nu_{H\alpha} \alpha^{eff}_{H\alpha}}} m_{p} n_{e}^{-1} \\
= 9.73 \times 10^{8} L_{H\alpha,43} n_{e,100}^{-1}\ {\rm M}_{\sun}
 \end{eqnarray}

Using $n_e$=500 cm$^{-3}$ and $L_{H\alpha,43}$=10$^{0.8}$, we find a
total mass necessary to sustain the H$\alpha$ luminosity of about 3
$\times$ 10$^{8}$ $M_{\sun}$. This is a rather modest amount of mass
and shows that it is very simple to have the AGN ionize this much gas,
which is only a small fraction of the total mass. 

Given these estimates, we appear to be able to explain the emission
line ratios with our estimated ionization parameter, looking at the
emission line diagrams and physical parameters for photoionized nebulae
in \cite{groves04, groves06}. The line ratios are indeed consistent
with having Hydrogen number densities of-order 100-1000 cm$^{-3}$ and
a dilute ionizing energy field (log $\bar{\rm U}_0\sim-3$).

There are of course analogs to this situation in both the local and high
redshift Universe. At high redshifts, optical emission line ionization
cones over scales of kpc to 100 kpc are seen in powerful radio galaxies
\citep[e.g.,][]{nesvadba06b,nesvadba08b}. Emission line nebulae this
large are generally seen in AGN with UV and emission line luminosities
that are higher than observed for BzK-15504 (about a factor of a few
to 10). However, the extended emission line region in BzK-15504 is
consistent with luminosities observed in the ``narrow line regions''
of distant QSOs \citep{netzer04}.

AGN also show strong asymmetries in their emission line
distributions. This may also explain the asymmetries seen in the emission
line images of BzK-15504 \citep{genzel06}. It represents the asymmetry
in the opening angle out of which the photons escape the AGN and where
the gas falls within the beam. A local example of this is NGC 1068
\citep{veilleux03}. In NGC 1068, the circum-nuclear ionization cone
shows a strong asymmetry in its distribution of \OIII$\lambda$5007
on both small scales \citep[100s of parsecs; ][]{evans93} and large
scale \cite[kpc scales; ][]{veilleux03}. Only in the areas of intense
H$\alpha$ emission are the line ratios consistent with photoionization
of massive stars in \HII\ regions.  Over a much larger scale, 10s of
kpc, in the areas of relatively low Hydrogen recombination line surface
brightness are the line ratios consistent with photoionization by the
AGN \citep{veilleux03}. Moreover, the emission line widths over the
regions excited by the AGN (FWHM$\sim$100-few 100 km s$^{-1}$) are also
in reasonable agreement with what is observed for BzK-15504. In many
optically selected AGN, the line widths are not strongly influenced
by the AGN despite the fact that the ionization state of the gas
is \citep{nelson96}. Of course, both the gas mass
and the power of the AGN are larger in BzK-15504 and it may well be
that as a result, the AGN can even outshine any star formation in the
extended optical emission line emitting gas (Fig.~\ref{fig:BPTNIIHa}
and \ref{fig:BPTSIIHaOIHa}). Within this regard it is also important
to note that BzK-15504 also has one of the highest H$\alpha$ surface
brightnesses in our sample.  The AGN could be responsible for increasing
its overall surface brightness as well as its AGN-like extended line
emission.  And like many classes of AGN (and in analogy with NGC1068 as
just discussed) the radiation, although likely heavily attenuated, is
able to escape to large distances despite the ionized gas representing
a small fraction of the total gas mass of BzK-15504.

This last point is important. AGN host galaxies show a
reasonable correlation between the size of their narrow
line region and the luminosity in H$\beta$ \citep[][but see,
\citealt{netzer04}]{bennert02}. The H$\beta$ luminosity of BzK-15504 would
suggest a narrow line region size of a few to 10 kpc (this is without
extinction correction). Interestingly, this is approximately the size
of the emission line nebulae observed in H$\alpha$ and [OIII].  Thus,
in agreement with other arguments, there is evidence that the extended
emission line region is nothing more than a narrow line region around
a powerful QSO with line widths that are, by analogy with local AGN,
influenced, perhaps dominated, by other processes.

\begin{figure*}
\plotone{fig9.eps}
\caption{The optical emission line ratios,
\SII$\lambda\lambda$6716,6731/H$\alpha$ versus
\OIII$\lambda$5007/H$\beta$ {\it (left)}, and \OI$\lambda$6300/H$\alpha$
versus \OIII$\lambda$5007/H$\beta$ {\it (right)} for the same subsample
as in Fig.~\ref{fig:BPTNIIHa}.  Again, the lines demarcate the regions
of the emission line diagrams where the photoionization is dominated by
star formation or AGN\citep{kewley01, osterbrock89}. These diagrams again
demonstrate the role of the AGN in photoionizing BzK-15504.  The other
galaxies in the sample lie close to the lines of demarcation in the
\SII$\lambda\lambda$6716,6731/H$\alpha$ versus \OIII$\lambda$5007/H$\beta$
diagram, but have limits that are consistent with being photoionized
mainly by massive stars.}
\label{fig:BPTSIIHaOIHa}
\end{figure*}

\section{Nature of the emission line gas in these
galaxies}\label{sec:EMGnature}

The surface brightness in our sample is both very high, and a function of
radius, velocity dispersion, and low ionization line emission ratio (i.e.,
\NII/H$\alpha$). This is in sharp contrast to what is observed for spiral
galaxies in the local Universe where the velocity dispersion is roughly
constant as the surface brightness declines exponentially.  With the
local trends in mind, we hypothesize that the trends seen in
the distant galaxy data can be explained by self-regulated star formation
whereby the intense star formation is pressure driven by the gas motions
induced by the star formation itself. We attempt to show that essentially,
these systems are analogous to the nearby starburst galaxies, such as M82
(and regulated perhaps in the same way as the ISM of the Milky Way) with an
overall similar surface brightness and similarly high pressure,
but with their star formation occurring over a much larger area.

\subsection{The relation between H$\alpha$ surface brightness and
\NII/H$\alpha$}\label{subsec:SBNIIoverHa}

In Fig.~\ref{fig:NIIHaSBHa}, we show the relationship between the emission
line ratio \NII/H$\alpha$ and the surface brightness of H$\alpha$. An
obvious way to understand such a relationship, in addition to the high
H$\alpha$ intensity, is through the gas pressure in the cloud or cloud
interfaces that gives rise to the recombination line emission.

To test this hypothesis we ran some simple photoionization models using the
code Cloudy\footnote{Calculations were performed with version 07.02.02 of
  Cloudy, last described by \citet{ferland98}.}. The input spectrum was
generated using Starburst99 \citep{leitherer99} for a constant star formation
rate and an age of 10$^8$ years \citep[consistent with the estimated ages of
  galaxies at z$\approx$2][]{erb06b}. However, the exact age is not very
important as long as it is sufficiently long for the spectral energy
distribution, especially in the UV, to reach an equilibrium shape. It is thus
appropriate for ages greater than about a few 10s Myrs and constant
star-formation rate. We assumed a constant density slab with ionization
parameters ranging from about log U=$-$5 to $-$1, initial densities log n=1,
2, and 3, and ISM metallicities and grain depletion which are kept constant
for all calculations. These attempts at modeling are not intended to be
exhaustive but illustrative. To gauge the impact of the ionization of the
modeled cloud, we ran these calculations for two total column densities,
10$^{20}$ and 10$^{21}$ cm$^{-2}$. The results from this modeling are shown
together with the data in Fig.~\ref{fig:NIIHaSBHa}.

We can see that the range of \NII/H$\alpha$ and surface brightness can
be explained by a combination of column density, volume densities, and
moderately diffuse radiation fields. Obviously, such simple assumptions
are not going to ``fit'' the data in any sense of the word, but show
that the area of line ratio-surface brightness space covered by the data
can be explained by a combination of these parameters, with each galaxy
exhibiting some range in each of these. Surface brightness is linearly
proportional to the density and ionization parameter, while the line ratio
depends on roughly one over the square root of the ionization parameter
for the range of ionization necessary to explain the line ratios. Such a
high density and ionization rate implies that the ISM of these galaxies
must in general exhibit high thermal pressures, P/k$\sim$10$^{6-7}$
K cm$^{-3}$ or more. The true maximum pressures are likely to be higher
as our physical resolution is only about a few kpc making it difficult
to identify the regions of highest surface brightness.

These pressures are much higher than observed in the disk of our
Milky Way or other nearby normal galaxies. As discussed in \citet{blitz06},
and references therein, the hydrostatic mid-plane pressure in the MW is
about 10$^{3.3-4.3}$ cm$^{-3}$ K, while for other local spirals it can
range up to about 10$^{6}$ cm$^{-3}$ K but is typically about 10$^{4.6}$
cm$^{-3}$ K. There have been fewer estimates of the more appropriate
comparison, namely the thermal pressures. In the nuclear regions of
local starburst galaxies, the thermal pressures as estimated from
the ratio \SII$\lambda$6716/\SII$\lambda$6731 give values that range
from 10$^{-9.3}$ to 10$^{-8.5}$ dyne cm$^{-2}$ or typically about
10$^7$ cm$^{-3}$ K \citep{lehnert96a} similar to that observed here at
z$\sim$2. Interestingly, \citet{brinchmann08} make a similar argument
for galaxies that generally fall off the locus of the \HII\ regions in
the \OIII/H$\beta$ versus \NII/H$\alpha$ diagram and lie close to the
dividing line between \HII\ region and AGN excitation. They find that
such galaxies have high specific star-formation rates, as do the general
population of high redshift galaxies discussed here, and relatively large
H$\alpha$ equivalent widths. \citeauthor{brinchmann08} suggest that the
most likely explanation for this is higher density (pressure) and that
the escape fraction may be higher in the star-forming regions implying
that the nebulae are at least partially density bounded. Similarly, in
Fig.~\ref{fig:NIIHaSBHa} we see that the clouds with column densities
of log N$_H$=20 become density bounded at high ionization parameters and
may explain the surface brightness limits we observe. Thus the pressures
estimated from the photoionization models are consistent with those of
nearby actively star-forming galaxies \citep{heckman90, wang99}.

\subsection{Consistency with high ionization lines}

In addition, for some of the galaxies we have H-band data
cubes of sufficient signal-to-noise to investigate the ratio of
\OIII$\lambda$5007 and H$\beta$. The photoionization models would predict
\OIII$\lambda$5007/H$\beta$ of about 0.5-2.0. Indeed, not considering
BzK-15504, which is powered by a QSO, the other galaxies show ratios
consistent with the high ionization line ratios and again amplifying
the idea that these nebula have increased density compared to nearby
galaxies as observed by \citet{brinchmann08}.

\section{Powering the local motions in these galaxies}

We have argued that the ISM of these distant galaxies is under high
pressure and that this high pressure could be induced by the intense
star formation within these systems. Fundamentally, this argument
is analogous to the situation in the most intense starbursts in the
local Universe such as M82 \citep[e.g.,][]{heckman90, lehnert96a,
lehnert99}. Such a hypothesis nicely explains some of the unusual
features in these distant star-forming galaxies such as their high surface
brightnesses in the recombination line(s) (which makes them observable
in the first place), the low ionization line ratios, the ratio of the
\SII\ lines (suggestive of high densities), and the trend for more
intense star formation to lead to broader lines. We observe low ratios
($\sim$2-4) of v/$\sigma$ in the extended line emitting regions of our
objects, much lower than the ratio of $\sim$10 generally observed in
local and intermediate redshift disk galaxies, and this appears to be
driven mainly by unusually high velocity dispersions, not low velocity
shears.

In principle, the high velocity dispersions observed in the ionized
gas lines are likely not representative of the turbulence in the
whole star-forming ISM. Nevertheless, the energy injection necessary
to sustain the high velocity dispersions observed in the ionized gas
should affect the atomic and molecular phases \citep[although if it is a
turbulent cascade, with energy injection on large scales, the velocity
dispersion will likely be lower in denser gas but the $\rho \sigma^2$
will be the same; ][]{joung08}. Indeed, high dispersions in the gas
out of which stars form is suggested by the large sizes and masses of
high-redshift star-forming regions \citep[giant clumps equaling the
Jeans mass, ][]{elmegreen05}. This is also supported by the thickness
of high-redshift disks when observed edge-on \citep[][]{EE06-thick}.

What is the power source of the high velocity dispersions that are
observed, and that should affect the whole ISM of high-redshift
galaxies? There are several possible mechanisms: (1) the conversion
of potential energy of infalling material in turbulent gas which
has sufficient angular momentum to relax into a disk configuration
\citep[cosmological accretion; ][]{fs06,dekel08}; (2) peculiar motions
in unstable disks that lead to perturbations in the velocity fields
that are not resolved in our data and lead to a high $\sigma$ without
consisting of real small-scale turbulence \citep[e.g.,][]{bournaud08};
(3) self-gravitationally generated turbulence \citep{wada02}; and/or
(4) the combined effect of the intense star formation.

To characterize the order-of-magnitude needed to power the turbulence,
we shall start by estimating its energy budget (if that is what these
motions represent).  Our low spatial resolution in the rest-frame of the
galaxies does not allow us to cleanly separate these possible sources of
the observed broad line emission. In fact, \citet{maclow99} as well as
our astrophysical reasoning below imply that bulk motion and turbulence
are intricately related in some of the scenarios, in the sense that bulk
motion will inject kinetic energy into the system which will then be
dissipated through turbulence. However, carefully examining the total
energy contained within the emission line gas may help us constrain
the mechanism responsible for its characteristics. We first consider
the energy dissipated within the turbulence and whether or not it is
consistent with cosmological accretion of gas.

\begin{figure}
\plotone{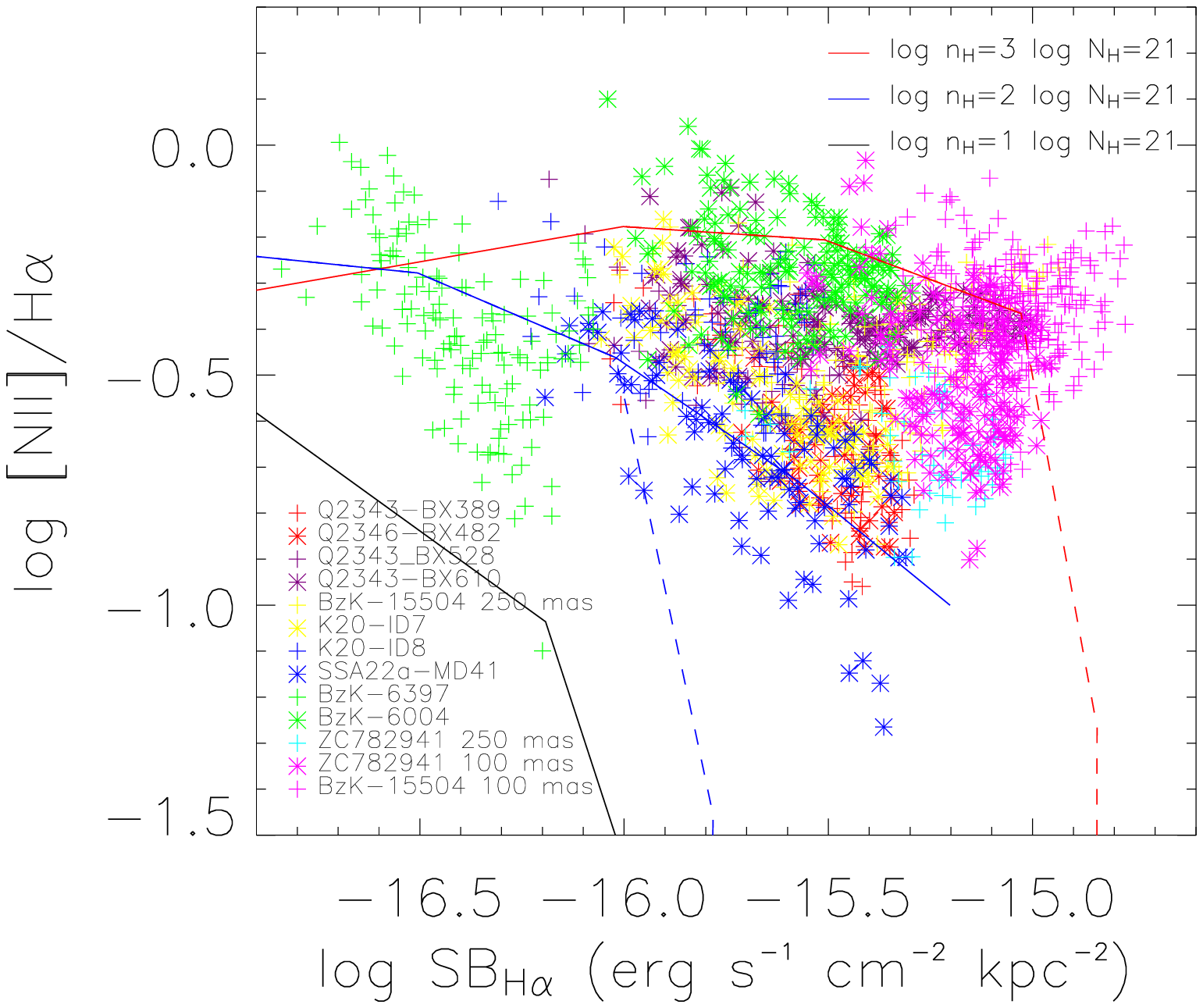}
\caption{The H$\alpha$ surface brightness (corrected
for surface brightness dimming) versus the logarithm of the
\NII$\lambda$6583/H$\alpha$ line ratio for all the individual spectra
of the galaxies in our sample. The galaxies are labeled in the legend to
the figure.  The lines represent results for photoionization modeling (see
text for details) for 6 set of conditions. The lines represent a range of
density (log n$_H$=1 to 3) and column densities of log N$_H$=21 cm$^{-2}$
(solid lines) and log N$_H$=20 cm$^{-2}$ (dotted lines). The ionization
parameters span from log U=$-$5 to $-$1.0 (which increases from left to
right along the lines, meaning low ionization parameters have relatively
high log \NII/H$\alpha$ and low H$\alpha$ surface brightnesses).}
\label{fig:NIIHaSBHa}
\end{figure}

\subsection{Turbulent energy dissipation}\label{subsec:Turbpower}

The violent motions observed in these distant galaxies are highly
supersonic given the densities derived previously for the optical emission
line gas and the turbulence generated, compressible.  To estimate the
dissipation of the turbulent energy therefore requires comparison with
(magneto-)hydrodynamic simulations of a realistic interstellar medium.
\citet{maclow99} provides an estimate of the energy dissipation rate for
compressible turbulence as,

\begin{equation}
\nonumber
\dot{\rm E}_{kin} \simeq -n_{\nu} m \tilde{k} v_{rms}^3
\end{equation}

where $n_{\nu}$ is a constant of proportionality, which \citet{maclow99}
estimates to be 0.21/$\pi$, $v_{rms}$ is the root mean square of
the velocity in the region, $m$ is the total mass, $E_{kin}$ is the
kinetic energy, and $\tilde{k}$ is the driving wavenumber.  Although,
the validity of this energy dissipation estimate on galaxy scales has
yet to be verified, it is useful to give an order-of-magnitude estimate
of the energy injection rate necessary to sustain the motions we observe.

The parameters necessary to estimate the energy dissipation rate of
compressible turbulence span a wide range of values. For example, the
velocity dispersions observed in the galaxies range from about $<$40 km
s$^{-1}$, in the outer regions, to about 250 km s$^{-1}$ in the inner,
circum-nuclear regions (Fig.~\ref{fig:HaSBsigma}). The typical velocity
dispersion is of-order $\sim$100-150 km s$^{-1}$.  The driving scale
is particularly difficult to estimate since the driving mechanisms
likely operate over a wide range of scales \citep[e.g.,][]{joung08}.
For example, if the largest scale driving mechanism was the cosmological
accretion of gas, we would expect this scale to be large \citep{dekel08}
approximately that of the thickness of the disk.  Alternatively, if the
driving mechanism is star-formation, the size of the largest star-forming
regions or associations in the galaxies might be the appropriate scale.
The clumps seen in Fig.~\ref{fig:nicmosimages} are approximately 100s
of parsecs to kpc in diameter. Since the driving wavenumber is inversely
proportional to the driving length, assuming a small scale for the driving
length would tend to increase the energy dissipated through turbulence.

We cannot estimate the total mass of gas in the galaxies in a straight
forward way. For simplicity, we will apply the Schmidt-Kennicutt law
\citep{kennicutt98a} to the star-formation intensities estimated from the
H$\alpha$ surface brightness distribution. This relation implies that
the gas surface densities are of-order $\Sigma_{gas} = 10^{2.6} {\rm
M}_{\sun} {\rm pc}^{-2}$ for star-formation intensities of 1 M$_{\sun}$
yr$^{-1}$ kpc$^{-2}$ (1 M$_{\sun}$ yr$^{-1}$ kpc$^{-2}$ is the typical
average value for the star-formation intensity approximately averaged
over the isophotal radius). This of course assumes that the molecular gas
has same kinematics as the warm ionized gas.  The high densities that we
found in \S~\ref{subsec:SBNIIoverHa} and the course angular resolution of
our data which averages the kinematics over a large region perhaps imply
that this is not a bad assumption \citep[but see ][]{joung08, walter02}.

If we adopt, $v_{rms}$=150 km s$^{-1}$, a mass surface density of 1000
M$_{\sun}$ pc$^{-2}$, a driving length of 1 kpc \citep[which is a typical
thickness of the ``clumpy disks'' observed at similar redshifts as our
sample;][]{EE06-thick}, we find that turbulence likely dissipates about
10$^{42}$ erg s$^{-1}$ kpc$^{-2}$.  The surface area within the isophotal
radius is approximately 200 kpc$^{2}$ (Table 1) giving a total energy
dissipation of about 10$^{44.3}$ erg s$^{-1}$.  We emphasize given
the uncertainties, the limits of our theoretical understanding of
turbulence and untested assumptions, this estimate of the turbulent
dissipation should be considered as order-of-magnitude only.

\subsection{Cosmological gas accretion}\label{subsec:cosacc}

We can compare this rough energy dissipation estimate with that of gas
infalling on to the disk itself. The total energy accretion rate from
gas falling onto the disk is given approximately by \citep{dekel07},

\begin{eqnarray}
\nonumber
\dot{\rm E}_{heating}=\mid \Delta \phi \mid \dot{M}_{gas}=4.8 \dot{M}_{gas} V_{c}^2 \\
= 10^{43.1}\dot{M}_{gas,100} V_{c,200}^2\ {\rm erg}\ {\rm s}^{-1}
\end{eqnarray}

where $\mid \Delta \phi \mid$ is the potential energy of infall,
$\dot{M}_{gas,100}$ is the halo gas accretion rate in units of 100
M$_{\sun}$ yr$^{-1}$ and $V_{c,200}$ is the circular velocity of the
halo in units of 200 km s$^{-1}$.  Thus it appears that accretion of
gas in itself cannot power the turbulent and bulk motions we observe
in these galaxies if these motions decay as compressable turbulence.
\citet{dekel09} also suggest that the main source of turbulence is not
infalling gas, unless this infalling gas is itself also highly clumpy,
something they themselves rule out as highly unlikely for a large fraction
of the infalling gas.

\subsection{Velocity dispersions in Jeans unstable clumps}

The dynamics of these galaxies could be influenced by the mode of
star formation. Elmegreen and collaborators have proposed that the
large number of {\it clump-cluster} and {\it chain galaxies} observed
at high resolution in deep HST imagery represent gas-rich Jeans
unstable disks \citep[review in][]{elmegreen-iau}. Clumpy galaxies
are not rare. Their frequency and relatively fast dynamical evolution
suggests that perhaps all galaxies pass through a ``clumpy'' stage in
their evolution and that this process is a natural way of explaining
phenomena like the growth of bulges and nuclear supermassive black holes
\citep{bournaud07b,elmegreen08}. HST/NICMOS images of five of our sample
galaxies (Fig.~\ref{fig:nicmosimages}) do show that they have ``clumpy''
irregular morphologies consistent with this hypothesis.

The clumpiness of the disks is hypothesized to be driven by
Jeans instability, which implies a relationship between the
mass of collapsing gas and the velocity dispersion within the gas
\citep[e.g.,][]{elmegreen07b}. Specifically, the Jeans relationship
implies that,

\begin{eqnarray}
\sigma_{gas}\sim M_J^{1/4} G^{1/2} \Sigma_{gas}^{1/4} = 54 M_{J,9}^{1/4} \Sigma_{SFR}^{0.18}\ {\rm km}\ {\rm s}^{-1}
\end{eqnarray}

where $\sigma_{gas}$ is the velocity dispersion of the gas, G is the
gravitational constant, $\Sigma_{gas}$ is the gas surface density in
M$_{\sun}$ pc$^{-2}$, $M_{J,9}$ is the Jeans mass in units of 10$^9$
M$_{\sun}$, and $\Sigma_{SFR}$ is the star-formation intensity in units
of M$_{\sun}$ yr$^{-1}$ kpc$^{-2}$. We have used the Schmidt-Kennicutt
relation \citep{kennicutt98b} to convert from gas surface density to
star-formation intensity. We show the relationship between the velocity
dispersion and star-formation intensity for Jeans unstable clumps
in Fig.~\ref{fig:SFIsigma} for a clump of 10$^9$ M$_{\sun}$, similar
to the largest masses estimated for clumps based on spectral energy
distribution fitting \citep{elmegreen05}.  We chose 10$^9$ M$_{\sun}$
to put the weakest limit on the possible contribution of the clumps to
the observed velocity dispersion.

The velocity dispersions predicted by Eq. (6) generally lie below the
data in Fig.~\ref{fig:SFIsigma}, especially at the highest intensities,
and the relationship is also too flat as a function of star-formation
intensity. Although this does not rule out clumps in a disk as a
contributing source to the high velocity dispersions observed in our
sample, especially for the gas with lowest dispersions, it cannot be the
entire explanation.  And taken at face value, the masses necessary to
explain the average dispersion would be more like 10$^{10}$ M$_{\sun}$,
not 10$^9$ M$_{\sun}$ suggesting that the dispersions are not dominated
by the internal dispersions of the clumps themselves.

\subsection{Gravity powering turbulence in dense gas-rich galaxies}

The clumpy light distributions observed suggest, in particular,
that the clumps themselves may have required an earlier source of
turbulence. This is necessary to ensure that the initial Jeans mass
is as high as the clumps masses observed \citep[up to more than $10^9
M_{\sun}$; ][]{elmegreen07b}. This implies that we may be obliged to
hypothesize another source of turbulence, especially in initiating the
intense star formation, to favor and encourage the growth of massive
clumps. Perhaps self-gravity powering the initial turbulence would play
the necessary role \citep[see][]{thomasson91,wada02}.

A gas disk with a low turbulent speed $\sigma$ may likely have a Toomre
parameter $Q = \frac{\sigma\kappa}{\pi G \Sigma} < 1$, if the surface
density $\Sigma$ is high. As a result, gravitational instabilities will
form and heat the gaseous medium, increasing its turbulent energy until
Q reaches $\sim 1$ and the process stops. The observation of clumps in
our z$\sim$2 galaxies and others suggest that Q is close to one, so that
this is the level at which gravity could indeed power the turbulence. In
local flocculent spiral galaxies, \citet{E03-flocculent} argued that
gravity was triggering the turbulence through local instabilities.

\begin{figure}
\epsscale{1.0}
\plotone{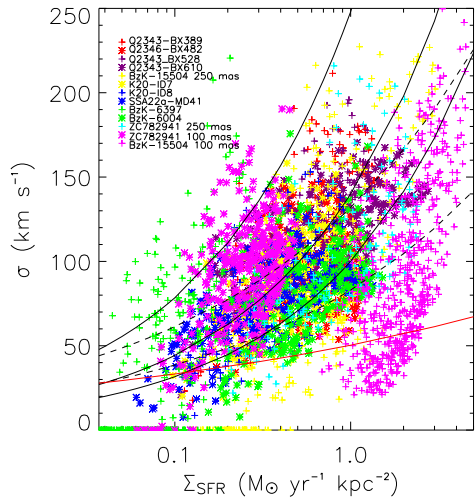}
\caption{A plot of the star-formation intensity versus the H$\alpha$
velocity dispersion in our sample of galaxies.  See Fig. 1 for details
on the 3$\times$3 pixel smoothing that was applied.  The symbols for the
different galaxies in our sample are shown in the legend to the figure.
The solid black lines show three simple relationship of the form
$\sigma=\surd{\epsilon\dot{E}}$, where $\epsilon$ is the efficiency of
coupling between the energy injected and the ISM.  {\it (from bottom
to top):} $\sigma=100 \surd\Sigma_{SFR}$ km s$^{-1}$, $\sigma=140
\surd\Sigma_{SFR}$ km s$^{-1}$, and $\sigma=240 \surd\Sigma_{SFR}$ km
s$^{-1}$ (see text for further details).  We note that the overall trend
seen in the diagram for the ensemble of galaxies is also traced by the
data from individual galaxies but with a somewhat steeper slope.  If the
dispersions represented turbulent motions, we would expect a scaling,
$\sigma=(\epsilon\dot{E})^{1/3}$. Using two scalings for the coupling
efficiency, 25\% and 100\% and a primary injection scale of 1 kpc, we
find $\sigma=80 \Sigma_{SFR}^{1/3}$ km s$^{-1}$ and $\sigma=130
\Sigma_{SFR}^{1/3}$ km s$^{-1}$ (bottom and top black dashed curves
respectively).  The red solid line represents the velocity dispersion
of a 10$^9$ M$_{\sun}$ clump using the simple Jeans relationship between
mass and velocity dispersion (see text for details). The offset between
galaxies may be due to a range of coupling efficiencies between the
mechanical energy output from the on-going star formation, differences
in the average pressure within the ISM, or geometrical factors such
as inclination.}
\label{fig:SFIsigma}
\end{figure}

The hydrodynamic simulations by \citet{agertz08} have shown that disk
self-gravity likely triggers the 5-10 km s$^{-1}$ turbulent speed
of extended HI disks around local spirals -- which could not result
from star formation beyond the edge of star-forming disks \citep[see
also][]{wada02}. The density of these modeled disks is nevertheless far
lower than the estimated density of gas disks at z$\sim$2. A gas fraction
of $\sim$50\% in these disks \citep[from][observations or inferred from
the Schmidt law, \citealt{bouche07}]{daddi08} indeed corresponds to
typical gas surface densities of $\sim$500 M$_{\sun}$ pc$^{-2}$. More
direct hints on the role of self-gravity in high-redshift disks can be
found in the models by \citet{tasker06,tasker08}, where gas disks with
large-scale densities around 100 M$_{\sun}$ pc$^{-2}$ are modeled, much
closer to the hypothesized density at z$\sim$2 even if still somewhat
lower. Interestingly, these models show larger clumps when the gas
density is large, indicating larger Jeans masses and hence larger velocity
dispersions. This is found in models of isolated disks without feedback,
where only gravity can power turbulence. The addition of feedback from
star-formation in these models does not change significantly the clump
masses, hence the turbulent speed, suggesting that the early stages of
turbulence development could be mostly powered by self-gravity. Direct
models of disks at z$\sim$2 with high density would be desirable, but
given that \citet{tasker08} have both the gas density and the clump
masses/sizes somewhat lower than z$\sim$2 standards, it is reasonable to
infer that disk self-gravity is likely important source of turbulence
at high redshift but does not likely generate the extreme dispersions
we observe (Fig.~\ref{fig:SFIsigma}).  As we noted earlier, it may be
that the formation of massive clumps themselves requires high turbulence
initially (Eqn. 5).

In addition, high velocity dispersions could be a result of mergers.
In the local Universe, intensely star-forming galaxies show very high
velocity dispersions in their optical emission line gas, up to 200
km s$^{-1}$ \citep{monreal-ibero06}.  To explain the low v/$\sigma$
observed at high redshift \citep{fs06}, \citet{robertson08} proposed that
a disk settling after a major gas-rich merger would have low v/$\sigma$.
In particular, they construct their model to match approximately the
properties of BzK-15504 in detail including its high star-formation rate
and intensity (without the AGN).  Thus, it is not clear how much of the
dispersion is due to the intense star-formation and how much is driven
purely by gravity in their model.  Moreover, recently \citet{bournaud09}
have criticised the merger model in that it does not naturally produce
the clumpy morphologies that are frequently observed in distant galaxies
and would lead to disks that make up a rather small fraction of the
total mass.  Thus it is not clear if gravitationally driven flows within
mergers can produce the dispersions and high emission line surface
brightness we observe without the accompanying intense star formation.

\subsection{H$\alpha$ surface brightness--velocity dispersion: Powering
the kinematics through star formation}\label{subsec:SBsigmarelation}

We argued above that self-gravity may be an important source of turbulence
at high redshift, but does not seem sufficient to generate the observed
velocity dispersions (see also Fig.~\ref{fig:SFIsigma}). This and the
relationship between H$\alpha$ surface brightness and velocity dispersion
suggests that the star formation within the galaxies is powering the
dynamics of the emission line gas. If the star formation is indeed
inducing the high pressures, then this is what would be expected. On
purely dimensional grounds, if the energy output from young stars is
controlling the dynamics of the emission line gas, we would expect,
that $\sigma$, the velocity dispersion, would be proportional to
the square root of the energy injection rate, dE$_{SF}$/dt, due to
stars. If the coupling efficiency of the mechanical energy output of
the star formation does not depend on radius, then the energy injection
rate is simply proportional to the star-formation rate. In this case
of course, the energy injection per unit area is proportional to the
star-formation intensity.  This hypothesis is equivalent to conserving
(with some efficiency) the mechanical energy output of the star-formation
within the ISM of the galaxy and that the velocities of the warm ionized
gas trace this energy injection rate.

In Fig.~\ref{fig:SFIsigma}, we have over-plotted just such a relationship.
This is not a fit to the data, but a simple scaling law based on the
star-formation rate per unit area and the velocity dispersion in the
warm neutral/ionized gas in the disk of the MW and other nearby galaxies.
The function is of the form, $\sigma$=($\epsilon \Sigma_{SFR}$)$^{1/2}$
where $\epsilon$ has been determined for the MW and other nearby galaxies
where the velocity dispersion in H$\alpha$ and star-formation intensities
have been related \citep[see][for details]{dib06}. There are several
possible values for $\epsilon$ in this relationship. If we take a simple
scaling from galaxies like the MW, they typically have star-formation
intensities in the regime of $\Sigma_{SFR}$=10$^{-5}$ to 10$^{-3}$
M$_{\sun}$ yr$^{-1}$ kpc$^{-2}$ and velocity dispersions in the warm
ionized gas of-order 10 km s$^{-1}$. \citeauthor{dib06} suggested
that galaxies may change from a quiescent disk mode to a starburst
mode at $\Sigma_{SFR}$=10$^{-2.5}$ to 10$^{-2}$ M$_{\sun}$ yr$^{-1}$
kpc$^{-2}$. This comes from modeling the ISM with a coupling efficiency
of 25\% to the supernova energy from disk star formation. Using these
two values for the scaling relation yields the bottom two curves in
Fig.~\ref{fig:SFIsigma}, whereas a coupling efficiency of 100\% would
yield the curve at the top in Fig.~\ref{fig:SFIsigma}. We emphasize
that the axis of abscissa is a simple scaling between H$\alpha$
surface brightness and the star-formation rate which was made assuming
a relationship between star-formation rate and H$\alpha$ luminosity
\citep{kennicutt98a}. We have not taken into account the effect of
extinction in making this scaling. If the galaxies for which we have
H$\beta$ measurements for are representative of our entire sample. This
would increase the star-formation intensities by a factor of a few
(Table 2).

While such a toy model is not a particularly good fit, especially at
the highest star-formation intensities, it has the virtue of having no
free parameters. It is just a simple scaling based on the MW and other
nearby galaxies. Although we put this in context of the modeling done
by \citet{dib06}, we could as well have simply used the MW scaling of
star-formation intensity and velocity dispersion. Obviously, the true
nature of the interstellar media of these distant galaxies cannot be so
simple as the energy injection argument presented here.  This argument
simply shows that the star formation in the galaxies themselves is likely
controlling the dynamics of the emission line gas.

\subsubsection{Mechanical energy due to star formation}\label{subsec:Lmech}

Given the relationship between star formation intensity and the
velocity dispersion of the gas, it is logical to investigate whether the
star formation has sufficient global energy injection rates to drive
the high velocity dispersions -- much like our previous estimate in
\S~\ref{subsec:Turbpower} comparing the accretion energy rate from infalling
cosmological gas.  Using the relationship from \cite{kennicutt98b}
between H$\alpha$ luminosity and star-formation rate, we find that
the total star-formation rate per unit area -- the star-formation
intensity -- ranges from about 0.05-5 M$_{\sun}$ yr$^{-1}$ kpc$^{-2}$
(Fig.~\ref{fig:HaSBsigma}). The dynamical time and integrated
H$\alpha$ equivalent width of our sample of galaxies and that from the
\citet{erb06b} sample, suggests that they may have been forming stars
continuously at the observed rate for the last few $\times$ 10$^8$ years.
This implies that the ongoing star formation in this sample in total
produces more than 10$^{41-42.5}$ erg s$^{-1}$ kpc$^{-2}$ of mechanical
energy \citep{leitherer99} for the range of large scale star-formation
intensities observed. These numbers are lower limits since we do not
have sufficient data to constrain the spatially resolved extinction and
so the true star-formation intensities are likely to be higher (Table 2).

We can now compare this total energy injection rate with the value
estimated \S~\ref{subsec:Turbpower}, namely 10$^{44.3}$ erg s$^{-1}$.  For a
fiducial value of 1 M$_{\sun}$ yr$^{-1}$ kpc$^{-2}$, which is typical
within a isophotal radius (Fig.~\ref{fig:HaSBsigma}), and the mechanical
energy output rate given above, would suggest total energy injection
rates of $>$10$^{44.2}$ erg s$^{-1}$.  Since the numbers are similar,
we suggest that the star-formation is powerful enough to maintain the
turbulent and bulk motions observed.

\subsubsection{Accelerating the emission line clouds}\label{subsec:acc}

Having now argued that the star formation is injecting energy into the
ISM of these galaxies and is sufficient to explain the overall dynamics
of the gas and to support dissipation through turbulence, can we make a
plausible model for how the clouds might be accelerated? \citet{klein94}
and \citet{nakamura06} have developed both analytic models and simulations
for clouds accelerated in a blast wave. Such a blast wave is expected
to be generated by the intense star formation observed in our sample
of galaxies. As the blast wave passes it shock heats the cloud and
eventually destroys it through Kelvin-Helmholtz and Rayleigh-Taylor
instabilities.  

In this scenario, the turbulence is driven by large scale bulk
motions induced by energy injection -- blast waves generated by intense
star-formation and other processes -- over a wide range of scales, which
then cascades into the denser gas and redistributes the energy over all
scales in the medium finally dissipating mainly on the smallest scales
\citep[][ and references therein]{joung08, maclow04}. The nature of
turbulence emphasizes the importance of both bulk and random motions and
estimating energy dissipation rates based on turbulence arguments while
estimating velocities based on bulk motion arguments are thus appropriate.

\citet{klein94} have found relatively simple
analytic approximations that can capture the destruction time of clouds
with a range of properties like contrast with the inter-cloud material,
Mach number of the blast wave, etc. Although there is a problem with
simply using these formulae with the parameters from the photoionization
calculations because those are aimed at providing the emission lines
without consideration of the shock heating in the first place, we envision
a scenario where shock heating, despite the intense energy input from
the supernova and stellar winds, is only a small fraction of the total
energy output of the star formation.  For our zeroth order model of
star formation proceeding for 10$^8$ years, we find that this condition
is clearly satisfied \citep{leitherer99}. Thus we expect the emission
lines to be dominated by the ionizing energy output from massive stars
and not by the shock heating.  The clouds could therefore be thought of
as the ones that have yet to be destroyed or are simply the surfaces of
sheets and filaments that will get eventually or are now being run over
by the blast waves into the ISM.

The relationship for the acceleration of the cloud is given by
\citet[][but see also \citealt{nakamura06}]{klein94}, as,

\begin{equation}
{\rm m_c} {{dv_c}\over{dt}} = -1/2 {\rm C_D} \rho_{i,1} \acute{v}_c^2 A
\end{equation}

where, ${\rm m_c}$ is the mass of the cloud, ${{dv_c}\over{dt}}$ is the
acceleration of the cloud, ${\rm C_D}\sim1$ is the drag coefficient,
$\rho_{i,1}$ is the post-shock density and is approximately, $\rho_{i,1}
\simeq 4\rho_{i,0}$ (we are assuming strong shocks), the pre-shock
density of the inter-cloud medium, and $\acute{v}_c = \mid v_{i,1}-v_c
\mid$ is the relative velocity of the shocked medium, where $v_{i,1}$
is the mean velocity of the cloud, and $A$ is the cross-sectional area.

Does the star formation actually generate blast waves of sufficient
energy to accelerate the clouds to the observed velocities? The total
mass ejected by supernovae and stellar winds in the star formation
process, $\dot{M}_{SF}$ can be scaled as $\dot{M}=\beta \dot{M}_{SF}$,
where $\beta$ is the fraction of the total ejected mass that is entrained
or mass loaded in the winds. Simple conservation arguments suggest the
terminal velocity of the outflow is v$_{\infty}$=$(\dot{E}/\dot{M})^{0.5}
\approx 2800 (\epsilon/\beta)^{0.5}$ km s$^{-1}$, where $\epsilon$
is the thermalization efficiency of the starburst mechanical energy
\citep[e.g.,][]{marcolini05, strickland09}. If we assume mass loading
rates of 1-10, then the blast wave speed is about 1000-2000 km
s$^{-1}$. Similarly, to first order the density of the blast wave is
proportional to $\beta^{3/2}$, which would suggest densities of-order
10$^{-3}$ cm$^{-3}$ \citep[e.g.,][]{marcolini05}. Such low densities and
high velocities meet the general criteria of being a blast wave, which
is highly supersonic for all of the densities in the ISM and should be
very efficient in destroying clouds as it passes through the ISM.

The modeling of the emission line gas in these galaxies suggest parameters
for the clouds like density and column density, and assuming the ISM
surrounding the pre-blast is like that typical of the ISM in local
galaxies, implies log $\rho_H$=$-$3 to 0, log $\rho_{cloud}$=1 to 3
cm$^{-3}$, r$_{cloud}$=0.1-10 pc. Using these values for the clouds, the
initial conditions for the ISM, and the blast wave formulation argued
for previously as input to the simple cloud acceleration models, we find
that the clouds could be accelerated up to 100 to a few 100 km s$^{-1}$
before being destroyed.  Thus it appears that at least in principle,
the velocities of the outflows generated by intense star formation
appear able to induce velocities like those observed. 

\subsection{Bulk and Turbulent motions}

Ultimately, following the above reasoning, star formation may give
rise to the observed line widths through a combination of bulk and
turbulent motion. This would also help resolve an apparent inconsistency
in our above arguments.  If turbulent motions dominate the observed
velocity dispersions, we may expect the relationship between the
dispersion and star-formation intensity of the form, $\sigma$=($\epsilon
\Sigma_{SFR}$)$^{1/3}$ (ignoring the relationship between star-formation
intensity and gas surface density which would tend to flatten this
relationship).  In Fig.~\ref{fig:SFIsigma}, we show this relationship
based on Eqn. 3.  To make this comparison with the data, we assume that
the energy of star-formation is dissipated entirely as turbulence, i.e.,
$\dot{E}_{kin}=\dot{E}_{SF}$ (see \S~\ref{subsec:Lmech}), assuming the
same parameters used in \S~\ref{subsec:Turbpower} and for two coupling
efficiencies between the energy injected by stars with the ISM.  We have
ignored the increase in the gas surface density with star-formation (the
Schmidt-Kennicutt relation).  While this relationship does explain the
overall trend in the data, data for individual galaxies are steeper than
$\sigma$=($\epsilon \Sigma_{SFR}$)$^{1/3}$ and in better agreement with
$\sigma$=($\epsilon \Sigma_{SFR}$)$^{1/2}$.  It could be this mixture
of bulk and turbulent velocities in the warm ionized gas steepens the
relationship between the dispersion and star-formation intensity from
$\sigma$=($\epsilon \Sigma_{SFR}$)$^{1/3}$.  But of course, as with
the scaling $\sigma$=($\epsilon \Sigma_{SFR}$)$^{1/2}$, this is too
simplistic.  One would need to consider the scale of energy injection,
what is the true nature of the warm ionized gas, the gas density, the
dependence of the distribution of the gas phases with star-formation
intensity, etc., in order to understand completely the underlying
mechanisms exciting the gas.

We emphasize that this analysis does not apply to all the emission
line gas, only the highest surface brightness gas. In nearby starburst
galaxies, very high velocity gas is observed, up to several hundred to
1000 km s$^{-1}$ \citep[e.g.,][]{heckman90, lehnert96a}. However, such
gas is generally of low surface brightness, well below the detection
limit of the data presented here \citep[e.g.,][]{heckman90,lehnert96a,
lehnert99}. As pointed out in, e.g., \citet{heckman90} and
\citet{lehnert96a}, the pressure in the emission line gas outside of
the intense star-forming regions in local starburst galaxies drops as
roughly radius$^{-2}$ and its surface brightness drops very rapidly
as well. Only the nuclear regions with extremely high pressures reach
surface brightnesses as observed in these distant galaxies and the
molecular gas in such regions can share similar outflow velocities as
the warm ionized gas \citep[e.g.,][ justifying our assumption that the
kinematics of the warm ionized and molecular gas may be similar on the
largest scales]{walter02}. We would therefore not expect to see the
highest velocity gas in H$\alpha$ or in the high and low ionization
lines observed \citep[see also][]{wang98, wang99}.

Turbulence is thought to be driven by large scale bulk motions induced
by energy injection -- blast waves generated by intense star-formation
and other processes -- over a wide range of scales, which then cascades
into the denser gas and redistributes the energy over all scales in
the medium finally dissipating mainly on the smallest scales \citep[][
and references therein]{joung08, maclow04}.  The nature of turbulence
emphasizes the importance of both bulk and random motions and estimating
energy dissipation rates based on turbulence arguments while estimating
velocities based on bulk motion arguments are perhaps appropriate.
Most likely, and in analogy with local starbursts, the warm ionized gas is
probing the interface between outflowing gas and dense clouds in the ISM
of these galaxies (especially at the highest surface brightnesses). It
is therefore probing the interface where bulk motion and thermal energy
is transfered to denser phases of gas through several mechanism such as
thermal instabilities, drag against the background flows, collisions
of cloud fragments, gas cooling from warm HII to HI to denser H$_2$
\citep[see ][ and references therein]{guillard09}.  Of course this
requires efficient conversion of bulk motions into turbulent energy which
is apparently the case \citep{maclow99}. The complexity of the processes
that control the dynamics and distribution of its various phases suggest
that no simple scaling like $\sigma$=($\epsilon \Sigma_{SFR}$)$^{1/2}$ can
provide an ultimate understanding of the ISM.  However, it does suggest
that the ISM of these distant galaxies is controlled by the intense
energy output of the star-formation within them.  Not surprising, but
something that needed to be investigated and certainly needs further study.

\section{Further Implications of Intense Star-formation}

It appears that it is the intense star formation (or perhaps to some
extent and in some cases the AGN) that is controlling the properties
of the emission line gas within these galaxies.  Given that we have
found that the star formation has sufficient mechanical energy output in
its own right to explain the characteristics of the emission line gas,
it is unclear whether the gas is telling us much about the underlying
mass distribution of these galaxies or their origins.  For example,
\citet{genzel08} have argued that high central velocity dispersions
require mass concentrations that are consistent with bulges.  We can now
explain this using the intense mechanical energy output of massive stars
(with a non-negligable contribution from AGN in some cases like BzK-15504)
and thus the velocity dispersions do not appear to be a unique tracer of
the underlying gravitational potential.  Similarly, it is plausible that
variations in the physical conditions of the gas will lead to variation
in the line ratios (see discussion in \S~\ref{subsec:SBNIIoverHa} and
Fig.~\ref{fig:NIIHaSBHa}).  Although these variations do not appear
sufficient to significantly affect the metallicity estimates based
on integrated spectra \citep[as argued by ][]{brinchmann08}, they are
sufficiently large that they may affect metallicity gradient estimates
within individual high redshift galaxies.

Determining that the ISM of these distant galaxies appears regulated
by the mechanical output of the intense star-formation, it is logical
to investigate how this output might influence the nature of the
star-formation itself and our understanding of the dynamics of
distant galaxies.  We do not as yet have a comprehensive theory
of star formation or a complete understanding of all the processes
that may limit both the efficiency and the rate at which stars can
form. Global star-formation laws or thresholds are those that describe
the gross characteristics of star formation on large scales and over
dynamical times of galaxies.  Several possibilities for explaining
the global characteristics of star formation have been suggested:
cloud-cloud collisions \citep[e.g.,][]{larson88, tan00} and the growth
of gravitational perturbations \citep{toomre64}, pressure and turbulence
regulated media \citep[e.g.,][]{silk97, silk01, elmegreen02, maclow04} and
others.  Models such as cloud-cloud collisions and pressure and turbulence
regulated ISM seem particularly appropriate in understanding the role
of self-regulation in intense star-formation.  However, of particular
interest in distant galaxies, and something that we can directly comment
on given our results, are the questions of the star-formation efficiency
and how does intense star-formation affect the structural properties
of distant galaxies.

\subsection{Efficiency of Star-Formation}

Recently, \citet{dekel08} investigated whether or not cosmological
gas accretion can sustain the star-formation rate of distant galaxies.
Galaxies with high star formation rates as observed here require the
star-formation rate to be close to the accretion rate of gas in this type
of model. This, on the face of it, requires the gas infall time scale and
the star-formation time scale to be close to equal which then implies
that the star-formation efficiency is high.

Our results suggest that the star-formation intensity, gas pressure,
and velocity dispersions in $z\sim2$ star-forming galaxies are similar or
higher than those in the most intense starbursting regions at $z=0$. The
efficiency of star-formation in local spirals is about a few percent,
but can reach $\sim 10$\% in starburst regions \citep[e.g.,][]{young96,
solomon-sage88, sanders91}. From this, we would infer efficiencies of
at least 10\% in our z$\sim$2 galaxies.  As mentioned previously, we do
not have a comprehensive theory of star formation.  However, it has been
suggested that there is a power-law dependence of the star formation rate
on the total column density \citep[][]{wang94}.  Such a relationship
implies that star-formation occurs on a dynamical time scale averaged
over a large area. \citet{elmegreen02} used this dependency to argue that
this would lead to a star-formation law of the form, $SFI=\epsilon_{SF}
\Sigma_{gas} \omega$, where $\epsilon_{SF}$ is the star-formation
efficiency and $\omega$ is a dynamical rate for the conversion of gas
into dense, star-forming cloud cores.  Interestingly, \citet{elmegreen02}
argue that this does not imply that gravitational forces are directly
involved but that this dynamical rate of conversion is about equal to
the turbulence crossing rate (which is also argued to be the inverse of
the collapse time with modest over-pressures).  Comparing the inverse
of the relative crossing times of turbulence of our sample and local
galaxies, $(v_{turb}/l)_{z\sim2}/(v_{turb}/l)_{z\sim0}\approx$2$-$5 for
$v_{turb}$=100$-$200 km s$^{-1}$ and l$\sim$1 kpc \citep{EE06-thick}
in the distant galaxies and $v_{turb}$=20$-$25 km s$^{-1}$ and
l$\sim$500 pc in local disks \citep{epinat09}. This suggests that
the conversion efficiency of gas into dense star-forming cloud cores
is higher in our sample of galaxies than for local disks.  Thus the
apparent efficiency of star-formation in our z$\sim$2 galaxies should be
higher than in nearby disk galaxies, where by ``apparent'' we mean the
product, $\epsilon_{SF}\omega$.  But it is important to note that the
star-formation efficiency itself, $\epsilon_{SF}$,  can remain constant.
On the other hand the efficiency can hardly exceed, say, 30\%, since
the star formation efficiency at the scale of individual cores is about
50\% \citep{matzner-mckee2000} and the efficiency in molecular clouds is
necessarily somewhat lower \citep{elmegreen02}.  Our observations thus
suggest ``apparent'' star-formation efficiencies between 10 and 30\%,
not just in a nuclear region but over the whole disk, which is globally
starbursting and highly turbulent.

An increase in the apparent efficiency of star-formation can be
understood with relation to the likely high surface density of gas
in these galaxies.  A high gas surface density is supported by both
the intense star-formation (through the Schmidt-Kennicutt relation)
and by the high pressures inferred from the optical emission line gas.
If this efficiency is related to the inverse of the gas consumption time
and the Schmidt-Kennicutt relation, then the gas suface density divided
by the star-formation intensity, the gas consumption time would be,
$\Sigma_{gas}/\Sigma_{SFR}\propto\Sigma_{gas}^{-0.5}$ -- a decreasing
function of the gas density.  Thus finding a high apparent efficiency
of star-formation is a natural consequence of having faster dynamical
processes such as faster collapse of gas due to high pressures, faster and
greater turbulent compression, and stronger influence of self-gravity.
Within this framework, it is easy to understand how the galaxies could
be clumpy and highly unstable against rapid and intense star-formation.

\subsection{Clumpy Disks and Dynamical Mass Estimates}
\label{ssec:implications}

Our results also have implications for the ``clumpy disk hypothesis''
\citep{elmegreen05,elmegreen07b}, whereby galaxy evolution is hypothesized
to be driven by internal clump formation \citep{bournaud07b}. Our
results suggest that the clumpy disk model alone cannot explain the
high dispersions observed and that an additional energy source is
needed. Gravity can play a crucial role, generating the high masses
observed in the clumps \citep{elmegreen05}, but we have argued that
star-formation must be an important source of energy too.  The intense
star-formation is a mechanism whereby further clump formation may be
stimulated by maintaining the high dispersions.  \citet{bournaud08}
conclude that the observations of the ``skywalker'' in the UDF (UDF6462)
and that the peculiar velocities observed are consistent with a
clumpy disk. Peculiar clump motions, typically around 50 km~s$^{-1}$
\citep{bournaud07b}, are likely masked by the turbulence and bulk
velocities generated in the optical emission line gas by the intense
star-formation. This implies that clump masses are difficult to infer
from the dispersions of their optical emission line gas. Furthermore,
while the clumpy disk model is certainly viable for a subset of the
observed galaxies, it is also likely that the mechanical output from the
intense star-formation within the clumps should not be ignored either
and may have a profound influence on the evolution of the clumps.

\citet{noguchi99} suggested that the bulges in local spiral galaxies
form through the dynamical evolution of massive clumps of stars
in high redshift galaxies. This model was further investigated by
\citet{elmegreen08}: large massive clumps migrate towards the disk center,
and coalesce into a slowly-rotating bulge. The limited observations
available may support this picture \citep{elmegreen09}. Could the high
efficiency of star formation in our observations of z$\sim$2 clumpy
galaxies question this model? Vigorous energy injection from intense
star formation may disrupt the clumps just like this process disrupts the
star-forming clouds in low redshift galaxies. Actually, if the stars are
forming in a higher density medium, which our observations may suggest,
the clumps may be more tightly bound and more difficult to destroy with
the mechanical energy output from massive stars \citep[even with strong
feedback from intense star formation, e.g., ][]{tasker08}. Therefore,
it is likely that the energy output affects mainly the lower-density,
inter-clump gas, but not the most massive clumps.  \citet{elmegreen08}
have independently argued that star-formation feedback could regulate
the bulge growth, but does not prevent this bulge-forming mechanism from
taking place in high redshift galaxies.

Finally, we have argued that the emission line ratios for BzK-15504 are
consistent with it harboring an AGN, and that this AGN is likely growing
very rapidly. In this situation, it is difficult to know how much of the
extended emission line gas is excited by star formation or the AGN and
what drives the dynamics of the extended gas. BzK-15504, since it has an
exquisite near-infrared data set, has been argued to be the archetypal
growing disk at high redshift. In fact, \citet{genzel08} have recently
proposed that this galaxy, and some others, represent a phase of rapid
gas accretion from their dark matter halo that is feeding its intense star
formation. If the extended emission line gas is influenced by a powerful
AGN, then this cannot be the case. The necessary gas masses and growth
rates would be lower, perhaps by factors of few or more, than previous
estimates if the AGN is increasing the overall surface brightness of the
extended emission line gas. We do not think that the AGN is the only
power source, but our data do not support the notion that this galaxy
must have an extreme accretion rate or a rapid rotation suggestive of a
massive disk. The ionization, surface brightness distribution, emission
line distribution asymmetry, and line widths are all consistent with
there being a kpc scale narrow line region surrounding a powerful QSO
in this object.

\section{Conclusions}

Our analysis of integral field spectroscopic data with SINFONI on the
ESO/VLT and HST/NICMOS imaging of our sample of z$\sim$2 galaxies have
revealed a number of interesting properties and relationships. Our results
suggest that the feedback cycle between the ISM and star formation (and
perhaps, more speculatively, AGN) plays an important, perhaps crucial
role in regulating star formation at high redshift.  In support of this
overall conclusion, we find:

\noindent
(1) These galaxies have surface brightnesses that are more than an
order of magnitude greater than those of local disks.  Our observations
suggest that the ISM of these z$\sim$ 2 galaxies is starbursting over
their whole area.

\noindent
(2)These galaxies have line ratios that suggest the presence of an AGN,
as in the case of BzK-15504, or high pressures and density bounded clouds.

\noindent
(3) Cosmological gas accretion models cannot provide sufficient energy
if its energy dissipates as compressible turbulence.

\noindent
(4) A relationship between the star-formation intensity and the velocity
dispersion of the emission line gas is found and it can be explained
by a simple energy injection relation. At low velocity dispersions,
self-gravity may play a role in generating turbulence.

\noindent
(5) The explanations for the relationships we observe imply that the
H$\alpha$ surface brightness and the distribution of velocity dispersion
may tell us little about the underlying mass distribution.  Thus bright
H$\alpha$ emission and broad lines are likely not evidence for mass
concentrations such as bulges or rings within these galaxies.

\noindent
(6) Given the high pressures and velocity dispersions we observe, it
is likely that the ISM (and perhaps star-formation) is turbulence and
pressure moderated \citep[e.g.,][]{silk97, silk01, elmegreen02, maclow04}.

\noindent
(7) Our results may imply apparent star-formation efficiencies 
as high as 10\% to 30\% (for the stellar mass formed per gas
mass unit per dynamical time).

Through several lines of argument, we outlined a picture in which
self-gravity generates high turbulence in gaseous disks.  Massive,
large and gas-rich clumps then form, triggering star formation with high
intensities ($>$0.1 M$_{\sun}$ kpc$^{-2}$) and apparent efficiencies
(10-30\%), with the ``starbursting'' interstellar medium spreading
over scales of 10-20 kpc. At this stage of intense star formation, the
turbulence and pressure moderated ISM likely regulates the star formation
through several possible mechanisms \citep[e.g.,][]{silk97, silk01,
elmegreen02, maclow04}.  The massive clumps however likely survive and
form bulges following the mechanisms proposed by \citet{noguchi99} and
discussed in further detail by \citet{elmegreen08}.  Such a picture has
a myriad of implications for our understanding of the most rapid periods
of star formation in distant galaxies and the ensemble properties of
galaxies generally.

\begin{acknowledgements} 

The authors wish to thank the staff at Paranal for their help and support
in making these observations.  MDL wishes to thank the Centre Nationale de
la Recherche Scientifique for its continuing support of his research and
NPHN wishes to acknowledge financial support from the European Commission
through a Marie Curie Postdoctoral Fellowship.  The work of ML, PDM,
and LL is directly supported by a grant from the Agence Nationale de
la Recherche (ANR). We acknowledge extensive discussions with Bruce
Elmegreen, Francois Boulanger, Avishai Dekel, and Emanuele Daddi and
Bruce Elmegreen specifically for his help and insight in understanding
the apparent star-formation efficiency.  We thank the referee for their
constructive and extremely helpful report that substantially improved
this manuscript.

\end{acknowledgements}

Facilities: \facility{ESO-VLT, HST(NICMOS)}

\clearpage 

\begin{deluxetable}{ccccccc}
\tabletypesize{\scriptsize}
\tablecaption{Properties of the High-z Galaxies\label{table:properties}}
\tablewidth{0pt}
\tablehead{\colhead{Object}&
\colhead{Line}&
\colhead{z}&
\colhead{FWHM}&
\colhead{flux}&
\colhead{SB limit}&
\colhead{r$_{iso}$}\\
\colhead{(1)}&\colhead{(2)}&\colhead{(3)}&\colhead{(4)}&\colhead{(5)}&\colhead{(
6)}&\colhead{(7)}}
\startdata
Q2343-BX610  &H$\alpha$&2.2098$\pm$0.0018&382$\pm$  8 &7.86$\pm$0.15&4.2&9.3\\
\nodata      &\NII\    &2.2104$\pm$0.0019&400$\pm$ 21 &3.23$\pm$0.15&\nodata&\nodata\\
\nodata      &\OIII\   &2.2090$\pm$0.0015&301$\pm$ 30 &0.67$\pm$0.06&\nodata&\nodata\\
\nodata      &H$\beta$ &2.2096$\pm$0.0019&383$\pm$ 62 &0.50$\pm$0.07&\nodata&\nodata\\
ZC782941$^a$ &H$\alpha$&2.1812$\pm$0.0015&305$\pm$ 13 &5.23$\pm$0.21&8.9&7.2\\
\nodata      &\NII\    &2.1816$\pm$0.0017&344$\pm$ 75 &1.15$\pm$0.22&\nodata&\nodata\\
ZC782941$^b$ &H$\alpha$&2.1814$\pm$0.0017&346$\pm$ 13 &3.05$\pm$0.11&3.8&4.2\\
\nodata      &\NII\    &2.1815$\pm$0.0018&378$\pm$ 42 &1.09$\pm$0.11&\nodata&\nodata\\
Q2343-BX528  &H$\alpha$&2.2684$\pm$0.0015&293$\pm$  8 &2.28$\pm$0.05&4.3&7.5\\
\nodata      &\NII\    &2.2689$\pm$0.0017&334$\pm$ 63 &0.43$\pm$0.06&\nodata&\nodata\\
BzK-15504$^b$&H$\alpha$&2.3816$\pm$0.0014&260$\pm$  8 &6.02$\pm$0.16&4.9&5.8\\
\nodata      &\NII\    &2.3818$\pm$0.0017&327$\pm$ 38 &2.54$\pm$0.22&\nodata&\nodata\\
BzK-15504    &H$\alpha$&2.3819$\pm$0.0020&400$\pm$ 12 &7.58$\pm$0.20&8.4&9.3\\
\nodata      &\NII\    &2.3813$\pm$0.0032&661$\pm$ 95 &3.34$\pm$0.38&\nodata&\nodata\\
\nodata      &\OIII\   &2.3833$\pm$0.0024&480$\pm$ 15 &8.68$\pm$0.24&\nodata&\nodata\\
\nodata      &H$\beta$ &2.3841$\pm$0.0012&182$\pm$ 50 &0.73$\pm$0.17&\nodata&\nodata\\
BzK-6397     &H$\alpha$&1.5132$\pm$0.0011&257$\pm$  8 &3.18$\pm$0.09&2.3&10.5\\
\nodata      &\NII\    &1.5135$\pm$0.0015&406$\pm$ 46 &1.08$\pm$0.11&\nodata&\nodata\\
BzK-6004K    &H$\alpha$&2.3865$\pm$0.0013&245$\pm$ 10 &4.28$\pm$0.16&7.7&8.7\\
\nodata      &\NII\    &2.3866$\pm$0.0015&274$\pm$ 22 &2.22$\pm$0.16&\nodata&\nodata\\
Q2346-BX482  &H$\alpha$&2.2563$\pm$0.0016&312$\pm$  6 &4.18$\pm$0.07&4.4&9.0\\
\nodata      &\NII\    &2.2564$\pm$0.0016&313$\pm$ 40 &0.64$\pm$0.07&\nodata&\nodata\\
\nodata      &\OIII\   &2.2569$\pm$0.0012&212$\pm$ 11 &1.62$\pm$0.07&\nodata&\nodata\\
\nodata      &H$\beta$ &2.2589$\pm$0.0011&166&0.22&\nodata&\nodata\\
K20-ID7      &H$\alpha$&2.2234$\pm$0.0016&313$\pm$  9 &4.02$\pm$0.10&5.3&9.3\\
\nodata      &\NII\    &2.2242$\pm$0.0014&282$\pm$ 50 &0.79$\pm$0.11&\nodata&\nodata\\
K20-ID8      &H$\alpha$&2.2231$\pm$0.0014&267$\pm$ 10 &2.51$\pm$0.09&7.2&7.5\\
\nodata      &\NII\    &2.2238$\pm$0.0014&264$\pm$ 50 &0.75$\pm$0.10&\nodata&\nodata\\
SSA22a-MD41  &H$\alpha$&2.1632$\pm$0.0015&298$\pm$ 12 &3.71$\pm$0.13&5.2&8.3\\
\nodata      &\NII\    &2.1627$\pm$0.0018&382$\pm$ 89 &0.72$\pm$0.16&\nodata&\nodata\\
\nodata      &\OIII\   &2.1704$\pm$0.0016&323$\pm$ 14 &2.29$\pm$0.09&\nodata&\nodata\\
\nodata      &H$\beta$ &2.1704$\pm$0.0009& 96$\pm$ 19 &0.44$\pm$0.07&\nodata&\nodata\\
Q2343-BX389  &H$\alpha$&2.1716$\pm$0.0024&525$\pm$ 18 &5.44$\pm$0.17&8.2&8.5\\
\nodata      &\NII\    &2.1732$\pm$0.0031&681$\pm$112 &1.65$\pm$0.22&\nodata&\nodata\\
\nodata      &\OIII\   &2.1713$\pm$0.0019&393$\pm$ 24 &1.65$\pm$0.09&\nodata&\nodata\\
\nodata      &H$\beta$ &2.1708$\pm$0.0010&141$\pm$ 28 &0.38$\pm$0.06&\nodata&\nodata \\
\enddata
\tablecomments{
Column (1) -- Object designation. $^a$ implies 250 mas pixel scale plus
adoptive optics, $^b$ implies 100 mas pixel scale plus adaptive optics.
No indication implies that the data were taken without the benefit of AO
and at 250 mas pixel$^{-1}$.
Column (2) -- Line identification.
Column (3) -- Redshift of the line in the integrated spectrum.  By integrated spectrum, we mean that the sum of the flux
from each pixel with a signal-to-noise greater than or equal to 3 in
each data cube for H$\alpha$ (see column 6 for this limiting value for H$\alpha$).  All of the sums for the other emission lines are over the same aperture
as for H$\alpha$.
Column (4) -- Full width at half maximum of the integrated spectrum of
each galaxy, corrected for instrumental resolution and is in units of
km s$^{-1}$.  
Column (5) -- Line flux of the integrated spectrum in units of 10$^{-16}$ erg s$^{-1}$ cm$^{-2}$.
Column (6) -- Surface brightness detection limit in units of 10$^{-19}$ erg s$^{-1}$ cm$^{-2}$ and defined at a signal to noise ratio, S/N$\approx$3.  This
is for a pixel that has been averaged over 3 pixels x 3 pixels.
Column (7) -- Isophotal radius defined as Area = $\pi$ r$_{\onehalf}^2$ where
the Area is defined as the projected area on the sky above a significance of
3 in the data and is in units of kpc.}
\end{deluxetable}

\clearpage

\begin{deluxetable}{ccccc}
\tablecaption{H$\beta$ Line Fluxes and Extinctions\label{table:extinctions}
}
\tablewidth{0pt}
\tablehead{\colhead{Source}
&\colhead{f$_{H\beta}$}
&\colhead{f$_{H\alpha}$/f$_{H\beta}$}
&\colhead{A$_{H\beta}$}
&\colhead{corr$_{H\alpha}$}\\
\colhead{(1)}&\colhead{(2)}&\colhead{(3)}&\colhead{(4)}&\colhead{(5)}}
\startdata
Q2343-BX610&0.61$\pm$0.08&8.1&2.8&4.6 \\
Q2343-BX389&0.37$\pm$0.05&4.1&1.0&1.7 \\
SSA22a-MD41&0.35$\pm$0.05&7.2&2.5&3.9 \\
BzK-15504  &0.52$\pm$0.08&7.1&2.4&3.8
\enddata
\tablecomments{
Column (1) -- Source designation.
Column (2) -- Measured line flux of H$\beta$ in units of 10$^{-16}$ erg s$^{-1}$ cm$^{-2}$ for the integrated spectrum.  In this case, we mean that the spectrum is integrated over the region where the S/N is greater than 3 in H$\beta$.
Column (3) -- Line ratio f$_{H\alpha}$/f$_{H\beta}$ for the integrated spectrum
as defined for column 2 in this table.
Column (4) -- The extinction in the H$\beta$ using the Galactic extinction law.
Column (5) -- The multiplicative factor required to correct H$\alpha$ for extinction assuming the Galactic extinction law.}
\end{deluxetable}

\begin{deluxetable}{ccccc}
\tablecaption{[SII] Emission Line Fluxes and Electron Densities\label{table:SII}
}
\tablewidth{0pt}
\tablehead{\colhead{Source}
&\colhead{f$_{\lambda6716}$}
&\colhead{f$_{\lambda6731}$}
&\colhead{f$_{\lambda6716}$/f$_{\lambda6731}$}
&\colhead{$n_{e}$} \\
\colhead{(1)}&\colhead{(2)}&\colhead{(3)}&\colhead{(4)}&\colhead{(5)}}
\startdata
Q2343-BX389 & 5.9$\pm$0.1      & 7.5$\pm$0.1      & 0.8$\pm$0.1  & 1200$^{+700}_{-400}$\\ 
Q2343-BX610 & 1.7$\pm$0.1      & 1.5$\pm$0.1      & 1.1$\pm$0.1  & 400$^{+700}_{-300}$ \\
Q2347-BX482 & 2.4$\pm$0.1      & 3.1$\pm$0.1      & 0.8$\pm$0.1  & 1200$^{+700}_{-400}$\\ 
BzK-15504   & 2.9$\pm$0.1      & 2.8$\pm$0.1      & 1.1$\pm$0.1  & 400$^{+700}_{-300}$ \\
BzK-6397    & 1.1$\pm$0.1      & 0.9$\pm$0.1      & 1.2$\pm$0.1  & 260$^{+150}_{-120}$
\enddata                                                                        \tablecomments{
Column (1) -- Source designation.  Column (2) -- Measured line flux of
\SII$\lambda$6716 in erg s$^{-1}$ cm$^{-2}$.  Column (3) -- Measured line
flux of \SII$\lambda$6731 in erg s$^{-1}$ cm$^{-2}$.  Column (4) -- Line
ratio F(6716)$/$F(6731).  Column (5) -- Electron density corresponding
to R$\pm 1\sigma$ for T$=10^4$ K in cm$^{-3}$.} 
\end{deluxetable}

\end{document}